\newif\ifuseTwoColumns
\ifuseTwoColumns
\documentclass[twocolumn]{IEEEtran}
\else
\documentclass[onecolumn,11pt,draftclsnofoot]{IEEEtran}
\usepackage{setspace}
\fi
\usepackage{amsmath}
\usepackage{amsfonts}
\usepackage{amssymb}
\usepackage{graphicx}
\usepackage{cases}
\usepackage{color}
\usepackage{mdwtab}
\usepackage{algorithm}
\usepackage{myalgorithmic}

\newcommand\T{\rule{0pt}{2ex}}
\newcommand\B{\rule[-1.2ex]{0pt}{0pt}}
\def\m(#1){\mathcal{#1}}
\def\b(#1){\mathbf{#1}}
\def\krank{\sigma}
\def\row(#1,#2){#1^{\textrm{row}}_#2}
\def\col(#1,#2){#1^{\textrm{col}}_#2}

\newcommand{\sinc}{\operatorname{sinc} }
\newcommand{\rank}{\operatorname{rank} }

\newcommand{\supp}{\operatorname{supp} }

\newcommand{\spark}{\operatorname{Spark} }
\def\colorname(#1){#1}
\newtheorem{theorem}{\colorname(Theorem)}
\newtheorem{lemma}{\colorname(Lemma)}

\newtheorem{proposition}{\colorname(Proposition)}

\newtheorem{definition}{\colorname(Definition)}

\title{
Blind Multi-Band Signal Reconstruction: Compressed Sensing for
Analog Signals
\thanks{The authors are with the Technion---Israel Institute of
Technology, Haifa Israel. Email: moshiko@tx.technion.ac.il,
yonina@ee.technion.ac.il.}}
\author{Moshe Mishali and Yonina C.~Eldar,
~\IEEEmembership{Member,~IEEE}}

\date{\today}
\usepackage[normalem]{ulem}

\begin{document}

\maketitle \IEEEpeerreviewmaketitle

\begin{abstract}
We address the problem of reconstructing a multi-band signal from
its sub-Nyquist point-wise samples. To date, all reconstruction
methods proposed for this class of signals assumed knowledge of
the band locations. In this paper, we develop a non-linear blind
perfect reconstruction scheme for multi-band signals which does
not require the band locations. Our approach assumes an existing
blind multi-coset sampling method. The sparse structure of
multi-band signals in the continuous frequency domain is used to
replace the continuous reconstruction with a single finite
dimensional problem without the need for discretization. The
resulting problem can be formulated within the framework of
compressed sensing, and thus can be solved efficiently using known
tractable algorithms from this emerging area. We also develop a
theoretical lower bound on the average sampling rate required for
blind signal reconstruction, which is twice the minimal rate of
known-spectrum recovery. Our method ensures perfect reconstruction
for a wide class of signals sampled at the minimal rate. Numerical
experiments are presented demonstrating blind sampling and
reconstruction with minimal sampling rate.
\end{abstract}


%%%%%%%%%%%%%%%%%%%%%%%%%%%%%%%%%%%%%%%%%%%%%%%%%%%%%%%%%
\section{Introduction}

\PARstart{T}{he} well known Whittaker, Kotel\'{n}ikov, and Shannon
(WKS) theorem links analog signals with a discrete representation,
allowing the transfer of the signal processing to a digital
framework. The theorem states that a real-valued signal
bandlimited to $B$ Hertz can be perfectly reconstructed from its
uniform samples if the sampling rate is at least $2B$ samples per
second. This minimal rate is called the Nyquist rate of the
signal.

Multi-band signals are bandlimited signals that posses an
additional structure in the frequency domain. The spectral support
of a multi-band signal is restricted to several continuous
intervals. Each of these intervals is called a band and it is
assumed that no information resides outside the bands. The design
of sampling and reconstruction systems for these signals involves
three major considerations. One is the sampling rate. The other is
the set of multi-band signals that the system can perfectly
reconstruct. The last one is blindness, namely a design that does
not assume knowledge of the band locations. Blindness is a
desirable property as signals with different band locations are
processed in the same way. Landau \cite{Landau} developed a
minimal sampling rate for an arbitrary sampling method that allows
perfect reconstruction. For multi-band signals, the Landau rate is
the sum of the band widths, which is below the corresponding
Nyquist rate.

Uniform sampling of a real bandpass signal with a total width of
$2B$ Hertz on both sides of the spectrum was studied in
\cite{Vaughan}. It was shown that only special cases of bandpass
signals can be perfectly reconstructed from their uniform samples
at the minimal rate of $2B$ samples/sec. Kohlenberg
\cite{Kohlenberg} suggested periodic non-uniform sampling with an
average sampling rate of $2B$. He also provided a reconstruction
scheme that recovers any bandpass signal exactly. Lin and
Vaidyanathan \cite{Vaidyanathan} extended his work to multi-band
signals. Their method ensures perfect reconstruction from periodic
non uniform sampling with an average sampling rate equal to the
Landau rate. Both of these works lack the blindness property as
the information about the band locations is used in the design of
both the sampling and the reconstruction stages.

Herley and Wong \cite{Herley} and Venkataramani and Bresler
\cite{Bresler00} suggested a blind multi-coset sampling strategy
that is called universal in \cite{Bresler00}. The authors of
\cite{Bresler00} also developed a detailed reconstruction scheme
for this sampling strategy, which is not blind as its design
requires information about the spectral support of the signal.
Blind multi-coset sampling renders the reconstruction applicable
to a wide set of multi-band signals but not to all of them.

Although spectrum-blind reconstruction was mentioned in two
conference papers in 1996 \cite{Bresler96_1D},\cite{Bresler96}, a
full spectrum-blind reconstruction scheme was not developed in
these papers. It appears that spectrum-blind reconstruction has
not been handled since then.

We begin by developing a lower bound on the minimal sampling rate
required for blind perfect reconstruction with arbitrary sampling
and reconstruction. As we show the lower bound is twice the Landau
rate and no more than the Nyquist rate. This result is based on
recent work of Lue and Do \cite{MinhDo} on sampling signals from a
union of subspaces.

The heart of this paper is the development of a spectrum-blind
reconstruction (SBR) scheme for multi-band signals. We assume a
blind multi-coset sampling satisfying the minimal rate
requirement. Theoretical tools are developed in order to transform
the continuous nature of the reconstruction problem into a finite
dimensional problem without any discretization. We then prove that
the solution can be obtained by finding the unique sparsest
solution matrix from Multiple-Measurement-Vectors (MMV). This set
of operations is grouped under a block we name \textit{Continuous
to Finite} (CTF). This block is the cornerstone of two SBR
algorithms we develop to reconstruct the signal. One is entitled
SBR4 and enables perfect reconstruction using only one instance of
the CTF block but requires twice the minimal sampling rate. The
other is referred to as SBR2 and allows for sampling at the
minimal rate, but involves a bi-section process and several uses
of the CTF block. Other differences between the algorithms are
also discussed. Both SBR4 and SBR2 can easily be implemented in
DSP processors or in software environments.

Our proposed reconstruction approach is applicable to a broad
class of multi-band signals. This class is the blind version of
the set of signals considered in \cite{Bresler00}. In particular,
we characterize a subset $\m(M)$ of this class by the maximal
number of bands and the width of the widest band. We then show how
to choose the parameters of the multi-coset stage so that perfect
reconstruction is possible for every signal in $\m(M)$. This
parameter selection is also valid for known-spectrum
reconstruction with half the sampling rate. The set $\m(M)$
represents a natural characterization of multi-band signals based
on their intrinsic parameters which are usually known in advance.
We prove that the SBR4 algorithm ensures perfect reconstruction
for all signals in $\m(M)$. The SBR2 approach works for almost all
signals in $\m(M)$ but may fail in some very special cases (which
typically will not occur). As our strategy is applicable also for
signals that do not lie in $\m(M)$, we present a nice feature of a
success recovery indication. Thus, if a signal cannot be recovered
this indication prevents further processing of invalid data.

The CTF block requires finding a sparsest solution matrix which is
an NP-hard problem \cite{Davis}. Several sub-optimal efficient
methods have been developed for this problem in the compressed
sensing (CS) literature \cite{Chen},\cite{Cotter}. In our
algorithms, any of these techniques can be used. Numerical
experiments on random constructions of multi-band signals show
that both SBR4 and SBR2 maintain a satisfactory exact recovery
rate when the average sampling rate approaches their theoretical
minimum rate requirement and sub-optimal implementations of the
CTF block are used. Moreover, the average runtime is shown to be
fast enough for practical usage.

Our work differs from other main stream CS papers in two aspects.
The first is that we aim to recover a continuous signal, while the
classical problem addressed in the CS literature is the recovery
of discrete and finite vectors. An adaptation of CS results to
continuous signals was also considered in a set of conferences
papers (see \cite{Analog2Info1},\cite{Analog2Info2} and the
references therein). However, these papers did not address the
case of multi-band signals. In \cite{Analog2Info2} an underlying
discrete model was assumed so that the signal is a linear
combination of a finite number of known functions. Here, there is
no discrete model as the signals are treated in a continuous
framework without any discretization. The second aspect is that we
assume a deterministic sampling stage and our theorems and results
do not involve any probability model. In contrast, the common
approach in compressive sensing assumes random sampling operators
and typical results are valid with some probability less than 1
\cite{Donoho},\cite{CandesRobust},\cite{Analog2Info1},\cite{Analog2Info2}.

The paper is organized as follows. In Section~\ref{Prelim} we
formulate our reconstruction problem. The minimal density theorem
for blind reconstruction is stated and proved in Section
\ref{SecMinRate}. A brief overview of multi-coset sampling is
presented in Section~\ref{SecOverview}. We develop our main
theoretical results on spectrum-blind reconstruction and present
the CTF block in Section~\ref{SecSBR}. Based on these results, in
Section~\ref{SecSBRAlg}, we design and compare the SBR4 and the
SBR2 algorithms. Numerical experiments are described in Section
\ref{SecNumerical}.

%%%%%%%%%%%%%%%%%%%%%%%%%%%%%%%%%%%%%%%%%%%%%%%%%%%%%%%%%%
\section{Preliminaries and Problem formulation}\label{Prelim}

%%%%%%%%%%%%%%%%%%%%%%%%%%%%%%%%%%%%%%%%%%%%%%%%%%%%%%%%%%
\subsection{Notation}

Common notation, as summarized in Table~\ref{TableNotation}, is
used throughout the paper. Exceptions to this notation are
indicated in the text.

\begin{table}[h] \caption{Notation}
\begin{tabular}[C]{| l | l |}
\hline
$x(t)$ & continuous time signal with finite energy\\
$X(f)$ & Fourier transform of $x(t)$ (that is assumed to
exist)\\
$a[n]$ & bounded energy sequence\\
$z^*$ & conjugate of the complex number $z$\\
$\b(v)$ & vector\\
$\b(v)_i$ or $\b(v)(i)$ & $i$th entry of $\b(v)$\\
$\b(v)(f)$ & vector that depends on a continuous parameter $f$\\
$\b(A)$ & matrix\\
$\b(A)_{ik}$ & $ik$th entry of $\b(A)$\\
$\b(A)^T,\b(A)^H$ & transpose and the conjugate-transpose of
$\b(A)$\\
$\b(A)\succeq 0$ & $\b(A)$ is an Hermitian positive semi-definite
(PSD) matrix
\\
$\b(A)^\dag$ & the Moore-Penrose pseudo-inverse of $\b(A)$\\
$S$ & finite or countable set\\
$S_i$ & $i$th element of $S$\\
$|S|$ & cardinality of a finite set $S$\\
$\m(T)$ & infinite non-countable set\\
$\lambda(\m(T))$ & the Lebesgue measure of $\m(T)\subseteq\mathbb{R}$\\
\hline
\end{tabular}
\label{TableNotation}
\end{table}

In addition, the following abbreviations are used.  The $\ell_p$
norm of a vector $\b(v)$ is defined as
\begin{displaymath}
\|\b(v)\|_p^p = \sum_i |\b(v)_i|^p,\qquad p\geq 0.
\end{displaymath}
The default value for $p$ is 2, so that $\|\b(v)\|$ denotes the
$\ell_2$ norm of $\b(v)$. The standard $L_2$ norm is used for
continuous signals. The $i$th column of $\b(A)$ is written as
$\b(A)_i$, the $i$th row is $(\b(A)^T)_i$ written as a column
vector.

Indicator sets for vectors and matrices are defined respectively
as
\begin{equation*}
I(\b(v)) = \{k\,|\,\b(v)(k)\neq 0\},\quad I(\b(A)) =
\{k\,|\,(\b(A)^T)_k \neq \b(0)\}.
\end{equation*}
The set $I(\b(v))$ contains the indices of non-zero values in the
vector $\b(v)$. The set $I(\b(A))$ contains the indices of the
non-identical zero rows of $\b(A)$.

Finally, $\b(A)_S$ is the matrix that contains the columns of
$\b(A)$ with indices belonging to the set $S$. The matrix
$\b(A)_S$ is referred to as the \textit{(columns) restriction} of
$\b(A)$ to $S$. Formally,
\begin{equation*}
(\b(A)_S)_i = (\b(A))_{S_i},\quad 1\leq i \leq |S|.
\end{equation*}
Similarly, $\b(A)^S$ is referred to as the \textit{rows
restriction} of $\b(A)$ to $S$.

%%%%%%%%%%%%%%%%%%%%%%%%%%%%%%%%%%%%%%%%%%%%%%%%%%%%%%%%%%
\subsection{Multi-band signals}\label{Prelim_MB}

\begin{figure}[h]
\centering
\includegraphics[scale=1]{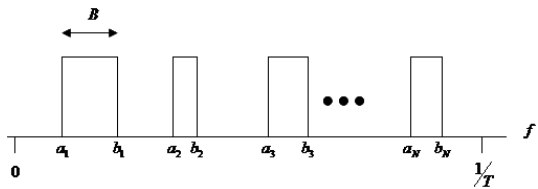}
\caption{Typical spectrum support of $x(t)\in\m(M)$.}
\label{TypicalSignal}
\end{figure}

In this work our prime focus is on the set $\m(M)$ of all
complex-valued multi-band signals bandlimited to $\m(F)=[0,1/T]$
with no more than $N$ bands where each of the band widths is upper
bounded by $B$. Fig.~\ref{TypicalSignal} depicts a typical
spectral support for $x(t)\in\m(M)$.

The Nyquist rate corresponding to any $x(t)\in\m(M)$ is $1/T$. The
Fourier transform of a multi-band signal has support on a finite
union of disjoint intervals in $\m(F)$. Each interval is called a
band and is uniquely represented by its edges $[a_i,b_i]$. Without
loss of generality it is assumed that the bands are not
overlapping.

Although our interest is mainly in signals $x(t)\in\m(M)$, our
results are applicable to a broader class of signals, as explained
in the relevant sections. In addition, the results of the paper
are easily adopted to real-valued signals supported on
$[-1/2T,+1/2T]$. The required modifications are explained in
Appendix~\ref{SecReal} and are based on the equations derived in
Section~\ref{SecMultiCoset}.

%%%%%%%%%%%%%%%%%%%%%%%%%%%%%%%%%%%%%%%%%%%%%%%%%%%%%%%%%%
\subsection{Problem formulation}\label{SecProbForm}

We wish to perfectly reconstruct $x(t)\in\m(M)$ from its
point-wise samples under two constraints. One is blindness, so
that the information about the band locations is not used while
acquiring the samples and neither can it be used in the
reconstruction process. The other is that the sampling rate
required to guarantee perfect reconstruction should be minimal.

This problem is solved if either of its constraints is removed.
Without the rate constraint, the WKS theorem allows perfect
blind-reconstruction for every signal $x(t)$ bandlimited to
$\m(F)$ from its uniform samples at the Nyquist rate $x(t=n/T)$.
Alternatively, if the exact number of bands and their locations
are known, then the method of \cite{Vaidyanathan} allows perfect
reconstruction for every multi-band signal at the minimal sampling
rate provided by Landau's theorem \cite{Landau}.

In this paper, we first develop the minimal sampling rate required
for blind reconstruction. We then use a multi-coset sampling
strategy to acquire the samples at an average sampling rate
satisfying the minimal requirement. The design of this sampling
method does not require knowledge of the band locations. We
provide a spectrum-blind reconstruction scheme for this sampling
strategy in the form of two different algorithms, named SBR4 and
SBR2. It is shown that if the sampling rate is twice the minimal
rate then algorithm SBR4 guarantees perfect reconstruction for
every $x(t)\in\m(M)$. The SBR2 algorithm requires the minimal
sampling rate and guarantees perfect reconstruction for most
signals in $\m(M)$. However, some special signals from $\m(M)$,
discussed in Section~\ref{SecSBR2}, cannot be perfectly
reconstructed by this approach. Excluding these special cases, our
proposed method satisfies both constraints of the problem
formulation.

%%%%%%%%%%%%%%%%%%%%%%%%%%%%%%%%%%%%%%%%%%%%%%%%%%%%%%%%%%
\section{Minimal sampling rate}\label{SecMinRate}

We begin by quoting Landau's theorem for the minimal sampling rate
of an arbitrary sampling method that allows known-spectrum perfect
reconstruction. It is then proved that blind
perfect-reconstruction requires a minimal sampling rate that is
twice the Landau rate.

%%%%%%%%%%%%%%%%%%%%%%%%%%%%%%%%%%%%%%%%%%%%%%%%%%%%%%%%%%
\subsection{Known spectrum support}

Consider the space of bandlimited functions restricted to a known
support  $\m(T) \subseteq \m(F)$:
\begin{equation}\label{KnownB_T}
\m(B)_\m(T)=\{x(t)\in L^2(\mathbb{R})\,|\,\supp X(f) \subseteq
\m(T)\}.
\end{equation}
A classical sampling scheme takes the values of $x(t)$ on a known
countable set of locations $R=\{r_n\}_{n=-\infty}^\infty$. The set
$R$ is called \textit{a sampling set} for $\m(B)_\m(T)$ if $x(t)$
can be perfectly reconstructed in a stable way from the sequence
of samples $x_R[n] = x(t=r_n)$. The stability constraint requires
the existence of constants $\alpha>0$ and $\beta<\infty$ such
that:
\begin{equation}\label{StabilitySubspaces}
\alpha\|x-y\|^2\leq \|x_R-y_R\|^2 \leq \beta \|x-y\|^2,\quad
\forall x,y \in \m(B)_\m(T).
\end{equation}
Landau \cite{Landau} proved that if $R$ is a sampling set for
$\m(B)_\m(T)$ then it must have a density $D^-(R)\geq
\lambda(\m(T))$, where
\begin{equation}\label{Breuling}
D^-(R) = \lim_{r\rightarrow \infty}\inf_{y\in
\mathbb{R}}\frac{|R\cap[y,y+r]|}{r}
\end{equation}
is the lower Beurling density, and $\lambda(\m(T))$ is the
Lebesgue measure of $\m(T)$. The numerator in (\ref{Breuling})
counts the number of points from $R$ in every interval of width
$r$ of the real axis\footnote{The numerator is not necessarily
finite but as the sampling set is countable the infimum takes on a
finite value.}. This result is usually interpreted as a minimal
\textit{average} sampling rate requirement for $\m(B)_\m(T)$, and
$\lambda(\m(T))$ is called the Landau rate.

%%%%%%%%%%%%%%%%%%%%%%%%%%%%%%%%%%%%%%%%%%%%%%%%%%%%%%%%%%
\subsection{Unknown spectrum support}\label{SecMinRateBlind}

Consider the set $\m(N)_\Omega$ of signals bandlimited to $\m(F)$
with bandwidth occupation no more than $0<\Omega< 1$, so that
\begin{equation*}
\lambda\left(\supp X(f)\right) \leq \frac{\Omega}{T},\quad\forall
x(t)\in\m(N)_\Omega.
\end{equation*}
The Nyquist rate for $\m(N)_\Omega$ is $1/T$. Note that
$\m(N)_\Omega$ is not a subspace so that the Landau theorem is not
valid here. Nevertheless, it is intuitive to argue that the
minimal sampling rate for $\m(N)_\Omega$ cannot be below
$\Omega/T$ as this value is the Landau rate had the spectrum
support been known.

A blind sampling set $R$ for $\m(N)_\Omega$ is a sampling set
whose design does not assume knowledge of $\supp X(f)$. Similarly
to (\ref{StabilitySubspaces}) the stability of $R$ requires the
existence of $\alpha>0$ and $\beta<\infty$ such that:
\begin{equation}\label{StabilitySubspacesBlind}
\alpha\|x-y\|^2\leq \|x_R-y_R\|^2 \leq \beta \|x-y\|^2,\quad
\forall x,y \in \m(N)_\Omega.
\end{equation}

\begin{theorem}[Minimal sampling rate]\label{BlindLandau}
Let $R$ be a blind sampling set for $\m(N)_\Omega$. Then,
\begin{equation}\label{MinimalDensity}
D^-(R)\geq \min\left\{\frac{2\Omega}{T},\frac{1}{T}\right\}.
\end{equation}
\end{theorem}

\begin{proof}
The set $\m(N)_\Omega$ is of the form
\begin{equation}
\m(N)_\Omega= \bigcup_{\m(T) \in \Gamma} \m(B)_\m(T),
\end{equation}
where
\begin{equation}
\Gamma = \{\m(T) \,|\, \m(T) \subseteq \m(F),\, \lambda(\m(T))
\leq \Omega/T\}.
\end{equation}
Clearly, $\m(N)_\Omega$ is a non-countable union of subspaces.
Sampling signals that lie in a union of subspaces has been
recently treated in \cite{MinhDo}. For every
$\gamma,\theta\in\Gamma$ define the subspaces
\begin{equation}
\m(B)_{\gamma,\theta}=\m(B)_\gamma+\m(B)_\theta = \{x+y\,|\,x\in
\m(B)_\gamma, y \in \m(B)_\theta\}.
\end{equation}
Since $R$ is a sampling set for $\m(N)_\Omega$,
(\ref{StabilitySubspacesBlind}) holds for some constants
$\alpha>0,\beta<\infty$. It was proved in \cite[Proposition
2]{MinhDo} that (\ref{StabilitySubspacesBlind}) is valid if and
only if
\begin{equation}\label{InvertSamplingSecant}
\alpha \|x-y\|^2 \leq \|x_R - y_R\|^2 \leq \beta \|x-y\|^2,\quad
\forall x,y\in \m(B)_{\gamma,\theta}
\end{equation}
holds for every $\gamma,\theta\in\Gamma$. In particular, $R$ is a
sampling set for every $\m(B)_{\gamma,\theta}$ with
$\gamma,\theta\in\Gamma$.

Observe that the space $\m(B)_{\gamma,\theta}$ is of the form
(\ref{KnownB_T}) with $\m(T)=\gamma \cup \theta$. Applying
Landau's density theorem for each $\gamma,\theta\in\Gamma$ results
in
\begin{equation}\label{MinimalDensityGammaTheta}
D^-(R)\geq \lambda(\gamma \cup
\theta),\quad\forall\gamma,\theta\in\Gamma.
\end{equation}
Choosing
\begin{equation*}
\gamma = \left[0,\frac{\Omega}{T}\right],\quad\theta =
\left[\frac{1-\Omega}{T},\frac{1}{T}\right],
\end{equation*}
we have that for $\Omega \leq 0.5$,
\begin{equation}\label{MinimalDensity1}
D^-(R)\geq \lambda(\gamma \cup \theta) = \lambda(\gamma) +
\lambda(\theta) = \frac{2\Omega}{T}.
\end{equation}
If $\Omega \geq 0.5$ then $\gamma\cup\theta=\m(F)$ and
\begin{equation}\label{MinimalDensity2}
D^-(R)\geq \lambda(\gamma\cup\theta)=\frac{1}{T}.
\end{equation}
Combining (\ref{MinimalDensity1}) and (\ref{MinimalDensity2})
completes the proof.
\end{proof}

In \cite{MinhDo}, the authors consider minimal sampling
requirements for a union of shift-invariant subspaces, with a
particular structure of sampling functions. Specifically, they
view the samples as inner products with sampling functions of the
form $\{\psi_k(t-m)\}_{1\leq k \leq K,m\in \mathbb{Z}}$, which
includes multi-coset sampling. Theorem~\ref{BlindLandau} extends
this result to an arbitrary point-wise sampling operator. In
particular, it is valid for non periodic sampling sets that are
not covered by \cite{MinhDo}.

An immediate corollary of Theorem~\ref{BlindLandau} is that if
$\Omega > 0.5$ then uniform sampling at the Nyquist rate with an
ideal low pass filter satisfies the requirements of our problem
formulation. Namely, both the sampling and the reconstruction do
not use the information about the band locations, and the sampling
rate is minimal according to Theorem~\ref{BlindLandau}. As $\m(M)$
is contained in the space of bandlimited signals, this choice also
provides perfect reconstruction for every $x(t)\in\m(M)$.
Therefore, in the sequel we assume that $\Omega\leq 0.5$ so that
the minimal sampling rate of Theorem~\ref{BlindLandau} is exactly
twice the Landau rate.

It is easy to see that $\m(M)\subset\m(N)_\Omega$ for
$\Omega=NBT$. Therefore, for known spectral support, the Landau
rate is $NB$. Despite the fact that $\m(M)$ is a true subset of
$\m(N)_{NBT}$, the proof of Theorem~\ref{BlindLandau} can be
adopted to show that a minimal density of $2NB$ is required so
that stable perfect reconstruction is possible for signals from
$\m(M)$.

We point out that both Landau's and Theorem~\ref{BlindLandau}
state a lower bound but do not provide a method to achieve the
bounds. The rest of the paper is devoted to developing a
reconstruction method that approaches the minimal sampling rate of
Theorem~\ref{BlindLandau}.

%%%%%%%%%%%%%%%%%%%%%%%%%%%%%%%%%%%%%%%%%%%%%%%%%%%%%%%%%%
\section{Universal Sampling}\label{SecOverview}

This section reviews multi-coset sampling which is used in our
development. We also briefly explain the fundamentals of
known-spectrum reconstruction as derived in \cite{Bresler00}.
%This section briefly reviews the multi-coset sampling for which
%SBR is developed. As pointed out earlier the parameter selection
%for this method is done irrespective of the specific band
%locations. Specific selection of the parameters is designated
%universal in \cite{Bresler00} as it ensures that the known
%spectrum reconstruction of \cite{Bresler00} applies to a wide
%class of signals. The conditions for universality are explained as
%they appear also in the development of the SBR scheme.

%%%%%%%%%%%%%%%%%%%%%%%%%%%%%%%%%%%%%%%%%%%%%%%%%%%%%%%%%%
\subsection{Multi-coset sampling}\label{SecMultiCoset}

Uniform sampling of $x(t)$ at the Nyquist rate results in samples
$x(t=nT)$ that contain all the information about $x(t)$.
Multi-coset sampling is a selection of certain samples from this
grid. The uniform grid is divided into blocks of $L$ consecutive
samples. A constant set $C$ of length $p$ describes the indices of
$p$ samples that are kept in each block while the rest are zeroed
out. The set $C=\{c_i\}_{i=1}^p$ is referred to as the sampling
pattern where
\begin{equation}
0\leq c_1 < c_2 < ... < c_p \leq L-1.
\end{equation}
Define the $i$th sampling sequence for $1\leq i \leq p $ as
\begin{numcases}{x_{c_i}[n]=}
\nonumber x(t=nT) & $n=mL+c_i,$ for some $m\in\mathbb{Z}$  \\
 0 & otherwise. \label{xci}
\end{numcases}
The sampling stage is implemented by $p$ uniform sampling
sequences with period $1/(LT)$, where the $i$th sampling sequence
is shifted by $c_i T$ from the origin. Therefore, a multi-coset
system is uniquely characterized by the parameters $L,p$ and the
sampling pattern $C$.

Direct calculations show that \cite{Bresler00}
\begin{align}\label{EqYAX}
X_{c_i}(e^{j2\pi
fT})=\frac{1}{LT}\sum\limits^{L-1}_{r=0}\exp\left(j\frac{2\pi}{L}
c_i r\right)X\left(f+\frac{r}{LT}\right),\\ \notag
\:\forall\,f\in\m(F)_0=\left[0,\frac{1}{LT}\right),\,1\leq i \leq
p,
\end{align}
where $X_{c_i}(e^{j2\pi fT})$ is the discrete-time Fourier
transform (DTFT) of $x_{c_i}[n]$. Thus, the goal is to choose
parameters $L,p,C$ such that $X(f)$ can be recovered from
(\ref{EqYAX}).

For our purposes it is convenient to express (\ref{EqYAX}) in a
matrix form as
\begin{equation}\label{My_yAx}
\b(y)(f)=\b(A)\b(x)(f),\; \; \forall f\in\m(F)_0,
\end{equation}
where $\b(y)(f)$ is a vector of length $p$ whose $i$th element is
$X_{c_i}(e^{j2\pi fT})$, and the vector $\b(x)(f)$ contains $L$
unknowns for each $f$
\begin{equation}\label{Xci}
\b(x)_i(f)=X\left(f+\frac{i}{LT}\right),\;\; 0\leq i\leq L-1,\;\;
f\in\m(F)_0.
\end{equation}
The matrix $\b(A)$ depends on the parameters $L,p$ and the set $C$
but not on $x(t)$ and is defined by
\begin{equation}\label{matrixA}
\b(A)_{ik} = \frac{1}{LT}\exp\left(j \frac{2\pi}{L} c_i k \right).
\end{equation}
%Note that $\b(A)$ is the rows restriction of the $L \times L$
%inverse-DFT matrix,
%\begin{equation}
%\b(W)_{ik} = \frac{1}{L}\exp\left(j \frac{2\pi}{L} i k \right),
%\end{equation}
%to the set $C$ with an additional constant of $1/T$.

Dealing with real-valued multi-band signals requires simple
modifications to (\ref{My_yAx}). These adjustments are detailed in
Appendix~\ref{SecReal}.

The Beurling lower density (i.e. the average sampling rate) of a
multi-coset sampling set is
\begin{equation}\label{MC_rate}
\frac{1}{T_\textrm{AVG}}=\frac{p}{LT},
\end{equation}
which is lower than the Nyquist rate for $p<L$. However, an
average sampling rate above the Landau rate is not sufficient for
known-spectrum reconstruction. Additional conditions are needed as
explained in the next section.

%%%%%%%%%%%%%%%%%%%%%%%%%%%%%%%%%%%%%%%%%%%%%%%%%%%%%%%%%%
\subsection{Known-spectrum reconstruction and universality}

The presentation of the reconstruction is simplified using CS
sparsity notation. A vector $\b(v)$ is called $K$-sparse if the
number of non-zero values in $\b(v)$ is no greater than $K$. Using
the $\ell_0$ pseudo-norm the sparsity of $\b(v)$ is expressed as
$\|\b(v)\|_0 \leq K$. We use the following definition of the
Kruskal-rank of a matrix \cite{Kruskal}:
\begin{definition}
The Kruskal-rank of $\b(A)$, denoted as $\sigma(\b(A))$, is the
maximal number $q$ such that every set of $q$ columns of $\b(A)$
is linearly independent.
\end{definition}

Observe that for every $f\in\m(F)_0$ the system of (\ref{My_yAx})
has less equations than unknowns. Therefore, a prior on $\b(x)(f)$
must be used to allow for recovery. In \cite{Bresler00} it is
assumed that the information about the band locations is available
in the reconstruction stage. This information supplies the set
$I(\b(x)(f))$ for every $f\in\m(F)_0$. Without any additional
prior the following condition is necessary for known-spectrum
perfect reconstruction
\begin{equation}\label{SparseVecXNec}
\b(x)(f)\textrm{ is $p$-sparse },\quad \forall f\in\m(F)_0.
\end{equation}
Using the Kruskal-rank of $\b(A)$ a sufficient condition is
formulated as
\begin{equation}\label{SparseVecXSuf}
\b(x)(f)\textrm{ is $\krank(\b(A))$-sparse },\quad \forall
f\in\m(F)_0.
\end{equation}
The known-spectrum reconstruction of \cite{Bresler00} basically
restricts the columns of $\b(A)$ to $I(\b(x)(f))$ and inverts the
resulting matrix in order to recover $\b(x)(f)$.

A sampling pattern $C$ that yields a fully Kruskal-rank $\b(A)$ is
called universal and corresponds to $\krank(\b(A))=p$. Therefore,
the set of signals that are consistent with (\ref{SparseVecXSuf})
is the broadest possible if a universal sampling pattern is used.
As we show later, choosing $L\leq \frac{1}{BT}$, $p\geq N$ and a
universal pattern $C$ makes (\ref{SparseVecXSuf}) valid for every
signal $x(t)\in\m(M)$.

Finding a universal pattern $C$, namely one that results in a
fully Kruskal-rank $\b(A)$, is a combinatorial process. Several
specific constructions of sampling patterns that are proved to be
universal are given in \cite{Bresler00},\cite{MishaliUniv}. In
particular, choosing $L$ to be prime renders every pattern
universal \cite{MishaliUniv}.

To summarize, choosing a universal pattern allows recovery of any
$x(t)$ satisfying (\ref{SparseVecXNec}) when the band locations
are known in the reconstruction. We next consider blind signal
recovery using universal sampling patterns.

%%%%%%%%%%%%%%%%%%%%%%%%%%%%%%%%%%%%%%%%%%%%%%%%%%%%%%%%%%
\section{Spectrum-Blind Reconstruction}\label{SecSBR}

In this section we develop the theory needed for SBR. These
results are then used in the next section to construct two
efficient algorithms for blind signal reconstruction.

The theoretical results are devoted in the following steps: We
first note that when considering blind-reconstruction, we cannot
use the prior of \cite{Bresler00}. In Section~\ref{SecUniqSBR} we
present a different prior that does not assume the information
about the band locations. Using this prior we develop a sufficient
condition for blind perfect reconstruction which is very similar
to (\ref{SparseVecXSuf}). Furthermore, we prove that under certain
conditions on $L,p,C$, perfect reconstruction is possible for
every signal in $\m(M)$. We then present the basic SBR paradigm in
Section~\ref{SecParadigm}. The main result of the paper is
transforming the continuous system of (\ref{My_yAx}) into a finite
dimensional problem without using discretization. In
Section~\ref{SecTransformMMV} we develop two propositions for this
purpose, and present the CTF block.

%%%%%%%%%%%%%%%%%%%%%%%%%%%%%%%%%%%%%%%%%%%%%%%%%%%%%%%%%%
\subsection{Conditions for blind perfect reconstruction}\label{SecUniqSBR}

Recall that for every $f\in\m(F)_0$ the system of (\ref{My_yAx})
is undetermined since there are fewer equations than unknowns. The
prior assumed in this paper is that for every $f\in\m(F)_0$ the
vector $\b(x)(f)$ is sparse but in contrast to \cite{Bresler00}
the location of the non-zero values is unknown. Clearly, in this
case (\ref{SparseVecXNec}) is still necessary for blind perfect
reconstruction. The following theorem from the CS literature is
used to provide a sufficient condition.

\begin{theorem}\label{SMVUniq}
Suppose $\b(\bar{x})$ is a solution of $\b(y)=\b(A)\b(x)$. If
$\|\b(\bar{x})\|_0\leq\krank(\b(A))/2$ then $\b(\bar{x})$ is the
unique sparsest solution of the system.
\end{theorem}
Theorem~\ref{SMVUniq} and its proof are given in \cite{MElad},
\cite{Chen} with a slightly different notation of $\spark(A)$
instead of the Kruskal-rank of $\b(A)$. Note that the condition of
the theorem is not necessary as there are examples that the
sparsest solution $\b(\bar{x})$ of $\b(y)=\b(A)\b(x)$ is unique
while $\b(\bar{x})>\krank(\b(A))/2$.

Using Theorem~\ref{SMVUniq}, it is evident that perfect
reconstruction is possible for every signal satisfying
\begin{equation}\label{SparseBlindCond}
\b(x)(f)\textrm{ is $\frac{\krank(\b(A))}{2}$ -sparse },\quad
\forall f\in\m(F)_0.
\end{equation}
As before, choosing a universal pattern makes the set of signals
that conform with (\ref{SparseBlindCond}) the widest possible.
Note that a factor of two distinguishes between the sufficient
conditions of (\ref{SparseVecXSuf}) and of
(\ref{SparseBlindCond}), and results from the fact that here we do
not know the locations of the non-zero values in $\b(x)(f)$.

Note that (\ref{SparseBlindCond}) provides a condition under which
perfect reconstruction is possible, however, it is still unclear
how to find the original signal. Although the problem is similar
to that described in the CS literature, here finding the unique
sparse vector must be solved for each value $f$ in the continuous
interval $\m(F)_0$, which clearly cannot be implemented.

In practice, conditions (\ref{SparseVecXSuf}) and
(\ref{SparseBlindCond}) are hard to verify since they require
knowledge of $x(t)$ and depend on the parameters of the
multi-coset sampling. We therefore prefer to develop conditions on
the class $\m(M)$ which characterizes multi-band signals based on
their intrinsic properties: the number of bands and their widths.
It is more likely to know the values of $N$ and $B$ in advance
than to know if the signals to be sampled satisfy
(\ref{SparseVecXSuf}) or (\ref{SparseBlindCond}). The following
theorem describes how to choose the parameters $L,p$ and $C$ so
that the sufficient conditions for perfect reconstruction hold
true for every $x(t)\in\m(M)$, namely it is a unique solution of
(\ref{My_yAx}). The theorem is valid for both known and blind
reconstruction with a slight difference resulting from the factor
of two in the sufficient conditions.

\begin{theorem}[Uniqueness]\label{SBR-Uniqueness}
Let $x(t) \in \m(M)$ be a multi-band signal. If:
\begin{enumerate}
\item The value of $L$ is limited by
\begin{equation}\label{LBT} L\leq \frac{1}{BT},
\end{equation}
\item $p\geq N$ for known reconstruction or $p\geq 2N$ for blind,
\item $C$ is a universal pattern,
\end{enumerate}
then, for every $f\in\m(F)_0$, the vector $\b(x)(f)$ is the unique
solution of (\ref{My_yAx}).
\end{theorem}

\begin{proof}
If $L$ is limited by (\ref{LBT}) then for the $i$th band
$\m(T)_i=[a_i,b_i]$ we have
\begin{equation*}
\lambda(\m(T)_i)\leq B \leq \frac{1}{LT}, \quad 1\leq i \leq N.
\end{equation*}
Therefore, $f\in \m(T)_i$ implies
\begin{equation*}
f + \frac{k}{LT}\notin \m(T)_i,\;\forall k\neq 0.
\end{equation*}
According to (\ref{Xci}) for every $f\in\m(F)_0$ the vector
$\b(x)(f)$ takes the values of $X(f)$ on a set of $L$ points
spaced by $1/LT$. Consequently, the number of non-zero values in
$\b(x)(f)$ is no greater than the number of the bands, namely
$\b(x)(f)$ is $N$-sparse.

Since $C$ is a universal pattern, $\krank(\b(A))=p$. This implies
that conditions (\ref{SparseVecXSuf}) and (\ref{SparseBlindCond})
are satisfied.
\end{proof}

Note that the condition on the value of $p$ implies the minimal
sampling rate requirement. To see this, substitute (\ref{LBT})
into (\ref{MC_rate}):
\begin{eqnarray}
\frac{1}{T_\textrm{AVG}} = \frac{p}{LT} \geq pB.
\end{eqnarray}
As pointed out in the end of Section~\ref{SecMinRateBlind}, if the
signals are known to lie in $\m(M)$ then the Landau rate is $NB$,
which is implied by $p\geq N$. Theorem~\ref{BlindLandau} requires
an average sampling rate of $2NB$, which can be guaranteed if $p
\geq 2N$.

%%%%%%%%%%%%%%%%%%%%%%%%%%%%%%%%%%%%%%%%%%%%%%%%%%%%%%%%%%
\subsection{Reconstruction paradigm}\label{SecParadigm}

The goal of our reconstruction scheme is to recover the signal
$x(t)$ from the set of sequences $x_{c_i}[n],\,\, 1\leq i\leq p$.
Equivalently, the aim is to reconstruct $\b(x)(f)$ of
(\ref{My_yAx}) for every $f\in\m(F)_0$ from the input data
$\b(y)(f)$.

A straight forward approach is to find the sparsest solution
$\b(x)(f)$ on a dense grid of $f\in\m(F)_0$. However, this
discretization strategy cannot guarantee perfect reconstruction.
In contrast, our approach is exact and does not require
discretization.

Our reconstruction paradigm is targeted at finding the diversity
set which depends on $x(t)$ and is defined as
\begin{equation}\label{setS}
S=\bigcup_{f\in\m(F)_0}I(\b(x)(f)).
\end{equation}
The SBR algorithms we develop in Section~\ref{SecSBRAlg} are aimed
at recovering the set $S$. With the knowledge of $S$ perfect
reconstruction of $\b(x)(f)$ is possible for every $f\in\m(F)_0$
by noting that (\ref{My_yAx}) can be written as
\begin{equation}\label{yAX_S}
\b(y)(f) = \b(A)_S \: \b(x)^S(f).
\end{equation}
If the diversity set of $x(t)$ satisfies
\begin{equation}\label{CondS}
|S|\leq\krank(\b(A)),
\end{equation}
then
\begin{equation}
(\b(A)_S)^\dag\b(A)_S = I,
\end{equation}
where $\b(A)_S$ is of size $p\times|S|$. Multiplying both sides of
(\ref{yAX_S}) by $(\b(A)_S)^\dag$ results in:
\begin{eqnarray}\label{ReconstructXf}
\b(x)^S(f) {=}\, (\b(A)_S)^\dag \b(y)(f),\quad\forall f\in\m(F)_0.
\end{eqnarray}
From (\ref{setS}),
\begin{equation}\label{ReconstructXfSc}
\b(x)_i(f)=0,\quad\forall f\in\m(F)_0,\,i\notin S.
\end{equation}
Thus, once $S$ is known, and as long as (\ref{CondS}) holds,
perfect reconstruction can be obtained by
(\ref{ReconstructXf})-(\ref{ReconstructXfSc}).

As we shall see later on (\ref{CondS}) is implied by the condition
required to transform the problem into a finite dimensional one.
Furthermore, the following proposition shows that for
$x(t)\in\m(M)$, (\ref{CondS}) is implied by the parameter
selection of Theorem~\ref{SBR-Uniqueness}.

\begin{proposition}\label{lemmaSizeS}
If $L$ is limited by (\ref{LBT}) then $|S|\leq 2N$. If in addition
$p\geq 2N$ and $C$ is universal then for every $x(t)\in\m(M)$, the
set $S$ satisfies (\ref{CondS}).
\end{proposition}

\ifuseTwoColumns
%% This figure is the CTF block
%%
\begin{figure*}
\centering
\includegraphics[scale=0.9]{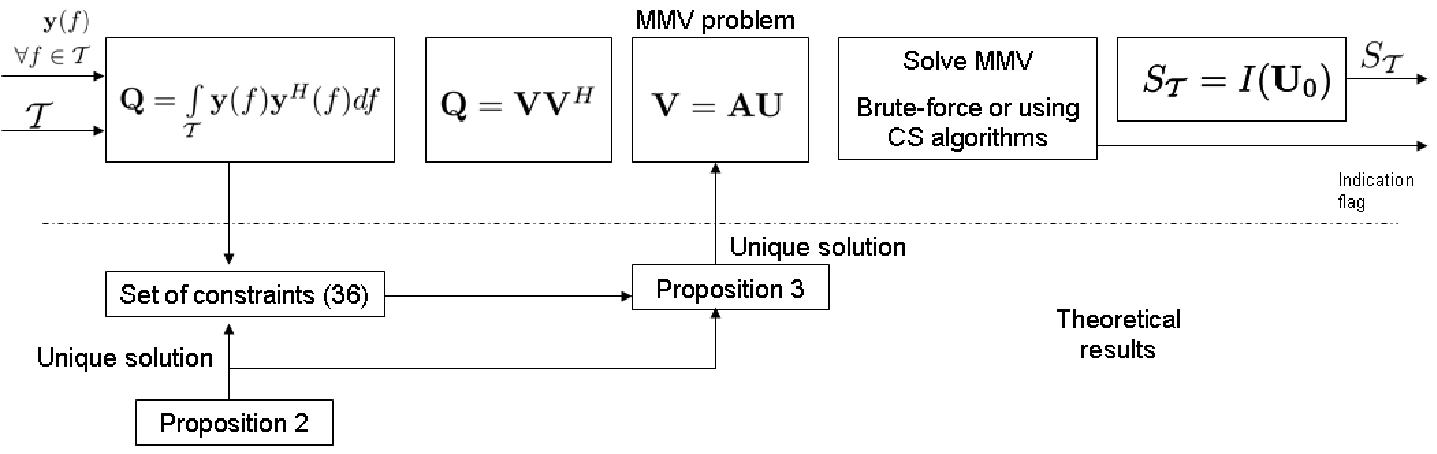}
\caption{Continuous to finite block (CTF). This block determines
the diversity set $\m(S)_\m(T)$ of a given interval $\m(T)$.}
\label{CTF}
\end{figure*}
\fi

\begin{proof}
The bands are continuous intervals upper bounded by $B$. From
(\ref{Xci}) it follows that $\b(x)(f)$ is constructed by dividing
$\m(F)$ into $L$ equal intervals of length $1/LT$. Therefore if
$L$ is limited by (\ref{LBT}) then each band can either be fully
contained in one of these intervals or it can be split between two
consecutive intervals. Since the number of bands is no more than
$N$ it follows that $|S|\leq 2N$. With the additional conditions
we have that $\krank(\b(A))=p\geq 2N \geq |S|$.
\end{proof}

As we described, our general strategy is to determine the
diversity set $S$ and then recover $x(t)$ via
(\ref{ReconstructXf})-(\ref{ReconstructXfSc}). In the non-blind
setting, $S$ is known, and therefore if it satisfies (\ref{CondS})
then the same equations can be used to recover $x(t)$. However,
note that when the band locations are known, we may use a value of
$p$ that is smaller than $2N$ since the sampling rate can be
reduced. Therefore, (\ref{CondS}) may not hold. Nonetheless, it is
shown in \cite{Bresler00}, that the frequency axis can be divided
into intervals such that this approach can be used over each
frequency interval. Therefore, once the set $S$ is recovered there
is no essential difference between known and blind reconstruction.

%%%%%%%%%%%%%%%%%%%%%%%%%%%%%%%%%%%%%%%%%%%%%%%%%%%%%%%%%%
\subsection{Formulation of a finite dimensional problem}\label{SecTransformMMV}

The set of equations of (\ref{My_yAx}) consists of an infinite
number of linear systems because of the continuous variable $f$.
Furthermore, the expression for the diversity set $S$ given in
(\ref{setS}) involves a union over the same continuous variable.
The main result of this paper is that $S$ can be recovered exactly
using only one finite dimensional problem. In this section we
develop the underlying theoretical results that are used for this
purpose.

Consider a given $\m(T)\subseteq\m(F)_0$. Multiplying each side of
(\ref{My_yAx}) by its conjugate transpose we have
%\begin{align}
%&\b(y)(f)\b(y)^H(f)=\left(\b(A)\b(x)(f)\right)\left(\b(A)\b(x)(f)\right)^H\\
%\nonumber &\phantom{\b(y)(f)\b(y)^H(f)} = \b(A)\b(x)(f) \b(x)^H(f)
%\b(A)^H,\quad \forall f\in\m(T).
%\end{align}
\begin{equation}
\b(y)(f)\b(y)^H(f)
= \b(A)\b(x)(f) \b(x)^H(f) \b(A)^H,\quad \forall
f\in\m(T).
\end{equation}
Integrating both sides over the continuous variable $f$ gives
\begin{equation}\label{QAZ} \b(Q)=\b(A)\b(Z)_0\b(A)^H,
\end{equation}
with the $p\times p$ matrix
\begin{equation}\label{MatQ} \b(Q) = \int_{f\in
\m(T)} \b(y)(f)\b(y)^H(f) df \succeq 0,
\end{equation}
and the $L\times L$ matrix
\begin{equation}\label{MatZ}
\b(Z)_0 = \int_{f\in \m(T)} \b(x)(f)\b(x)^H(f) df \succeq 0.
\end{equation}
Define the diversity set of the interval $\m(T)$ as
\begin{equation}\label{setST}
S_\m(T) = \bigcup_{f\in\m(T)}I(\b(x)(f)).
\end{equation}
Now,
\begin{equation*}
(\b(Z)_0)_{ii} = \int_{f\in \m(T)} |\b(x)_i(f)|^2 df.
\end{equation*}
This means that $(\b(Z)_0)_{ii}=0$ if and only if
$\b(x)_i(f)=0,\forall f\in\m(T)$, which implies that
$S_\m(T)=I(\b(Z)_0)$.

The next proposition is used to determine whether $\b(Z)_0$ can be
found by a finite dimensional problem. The proposition is stated
for general matrices $\b(Q),\b(A)$.

\begin{proposition}\label{TheoremFinite}
Suppose $\b(Q)\succeq 0$ of size $p\times p$ and $\b(A)$ are given
matrices. Let $\b(Z)$ be any $L\times L$ matrix satisfying
\begin{subequations}
\begin{equation}\label{CondZCon1}
\b(Q)=\b(A)\b(Z)\b(A)^H,
\end{equation}
\begin{equation}\label{CondZCon2}
\b(Z)\succeq 0,
\end{equation}
\begin{equation}\label{CondZCon3}
|I(\b(Z))|\leq \krank(\b(A)).
\end{equation}
Then, $\rank(\b(Z))=\rank(\b(Q))$. If, in addition,
\begin{equation}\label{CondZCon4}
|I(\b(Z))| \leq \frac{\krank(\b(A))}{2},
\end{equation}
\end{subequations}
then, $\b(Z)$ is the unique solution of
(\ref{CondZCon1})-(\ref{CondZCon4}).
\end{proposition}

\begin{proof}
Let $\b(Z)$ satisfy (\ref{CondZCon1})-(\ref{CondZCon3}). Define
$r_Q = \rank(\b(Q)),\,r_Z = \rank(\b(Z))$. Since $\b(Z)\succeq 0$
it can be decomposed as $\b(Z)=\b(P)\b(P)^H$ with $\b(P)$ of size
$L\times r_Z$ having orthogonal columns. From (\ref{CondZCon1}),
\begin{equation}\label{QAP}
\b(Q)=(\b(A)\b(P))(\b(A)\b(P))^H.
\end{equation}
It can be easily be concluded that $I(\b(Z))=I(\b(P))$, and thus
$|I(\b(P))|\leq\krank(\b(A))$. The following lemma whose proof is
given in Appendix~\ref{ProofLemIn} ensures that the matrix
$\b(A)\b(P)$ of size $p\times r_Z$ also has full column rank.

\begin{lemma}\label{LemIn}
For every two matrices $\b(A),\b(P)$, if $|I(\b(P))|\leq
\krank(\b(A))$ then $\rank(\b(P)) = \rank(\b(A)\b(P))$.
\end{lemma}

Since for every matrix $\b(M)$ it is true that
$\rank(\b(M))=\rank(\b(M)\b(M)^H)$, (\ref{QAP}) implies $r_Z=r_Q$.
%\bibitem{HornJohnsonBook}
%R.~A.~Horn and C.~R.~Johnson,
%\newblock {\em Matrix Analysis.}
%\newblock New York: Cambridge Univ. Press, 1991.
%According to \cite[Observation 7.1.6]{HornJohnsonBook} every
%matrix $\b(M)$ satisfies
%\begin{equation*}
%\rank(\b(M)) = \rank(\b(M)\b(M)^H).
%\end{equation*}
%Therefore Eq. (\ref{QAP}) implies $r_Z=r_Q$.

For the second part of Proposition~\ref{TheoremFinite} suppose
that $\b(Z),\b(\tilde{Z})$ both satisfy
(\ref{CondZCon1}),(\ref{CondZCon2}),(\ref{CondZCon4}). From the
first part,
\begin{equation*}
\rank(\b(Z)) = \rank(\b(\tilde{Z})) = r_Q.
\end{equation*}
Following the earlier decompositions we write
\begin{eqnarray}\label{ZZtilde}
\b(Z)=\b(P)\b(P)^H, \quad I(\b(Z))=I(\b(P)) \\ \nonumber
\b(\tilde{Z})=\b(\tilde{P})\b(\tilde{P})^H , \quad
I(\b(\tilde{Z}))=I(\b(\tilde{P})).
\end{eqnarray}
In addition,
\begin{equation}\label{UU2N}
|I(\b(P))|\leq \frac{\krank(\b(A))}{2}, \quad
|I(\b(\tilde{P}))|\leq \frac{\krank(\b(A))}{2}.
\end{equation}

From (\ref{CondZCon1}),
\begin{eqnarray}\label{AUAUAUAU}
\b(Q) = (\b(A)\b(P))(\b(A)\b(P))^H =
(\b(A)\b(\tilde{P}))(\b(A)\b(\tilde{P}))^H,
\end{eqnarray}
which implies that
\begin{equation}\label{APPR}
\b(A)(\b(P)-\b(\tilde{P})\b(R))=0,
\end{equation}
for some unitary matrix $\b(R)$. It is easy to see that
(\ref{UU2N}) results in $|I(\b(\tilde{P})\b(R))|\leq
\krank(\b(A))/2$. Therefore the matrix $\b(P)-\b(\tilde{P})\b(R)$
has at most $\krank(\b(A))$ non-identical zero rows. Applying
Lemma~\ref{LemIn} to (\ref{APPR}) results in
$\b(P)=\b(\tilde{P})\b(R)$. Substituting this into (\ref{ZZtilde})
we have that $\b(Z)=\b(\tilde{Z})$.
\end{proof}

The following proposition shows how to construct the matrix
$\b(Z)$ by finding the sparsest solution of a linear system.

\begin{proposition}\label{PropMMV}
Consider the setting of Proposition~\ref{TheoremFinite} and assume
$\b(Z)$ satisfies (\ref{CondZCon4}). Let $r=\rank(\b(Q))$ and
define a matrix $\b(V)$ of size $p \times r$ using the
decomposition $\b(Q)=\b(V)\b(V)^H$, such that $\b(V)$ has $r$
orthogonal columns. Then the linear system
\begin{equation}\label{VAU}
\b(V)=\b(A)\b(U)
\end{equation}
has a unique sparsest solution matrix $\b(U)_0$. Namely,
$\b(V)=\b(A)\b(U)_0$ and $|I(\b(U)_0)|$ is minimal. Moreover,
$\b(Z)=\b(U)_0\b(U)^H_0$.
\end{proposition}

\begin{proof}
Substitute the decomposition $\b(Q)=\b(V)\b(V)^H$ into
(\ref{CondZCon1}) and let $\b(Z)=\b(P)\b(P)^H$. The result is
$\b(V)=\b(A)\b(P)\b(R)$ for some unitary $\b(R)$. Therefore, the
linear system of (\ref{VAU}) has a solution $\b(U)_0 =
\b(P)\b(R)$. It is easy to see that
$I(\b(U)_0)=I(\b(P))=I(\b(Z))$, thus (\ref{CondZCon4}) results in
$|I(\b(U)_0)|\leq\krank(\b(A))/2$. Applying Theorem~\ref{SMVUniq}
to each of the columns of $\b(U)_0$ provides the uniqueness of
$\b(U)_0$. It is trivial that $\b(Z)=\b(U)_0\b(U)^H_0$.
\end{proof}

Using the same arguments as in the proof it is easy to conclude
that $I(\b(Z))=I(\b(U)_0)$, so that $S_\m(T)$ can be found
directly from the solution matrix $\b(U)_0$. In particular, we
develop the \textit{Continuous to Finite} (CTF) block which
determines the diversity set $S_\m(T)$ of a given frequency
interval $\m(T)$. Fig.~\ref{CTF} presents the CTF block that
contains the flow of transforming the continuous linear system of
(\ref{My_yAx}) on the interval $\m(T)$ into the finite dimensional
problem of (\ref{VAU}) and then to the recovery of $S_\m(T)$. The
role of Propositions~\ref{TheoremFinite}~and~\ref{PropMMV} is also
illustrated. The CTF block is the heart of the SBR scheme which we
discuss next.

\ifuseTwoColumns\else
%% This figure is the CTF block
%%
\begin{figure*}
\centering
\includegraphics[scale=0.9]{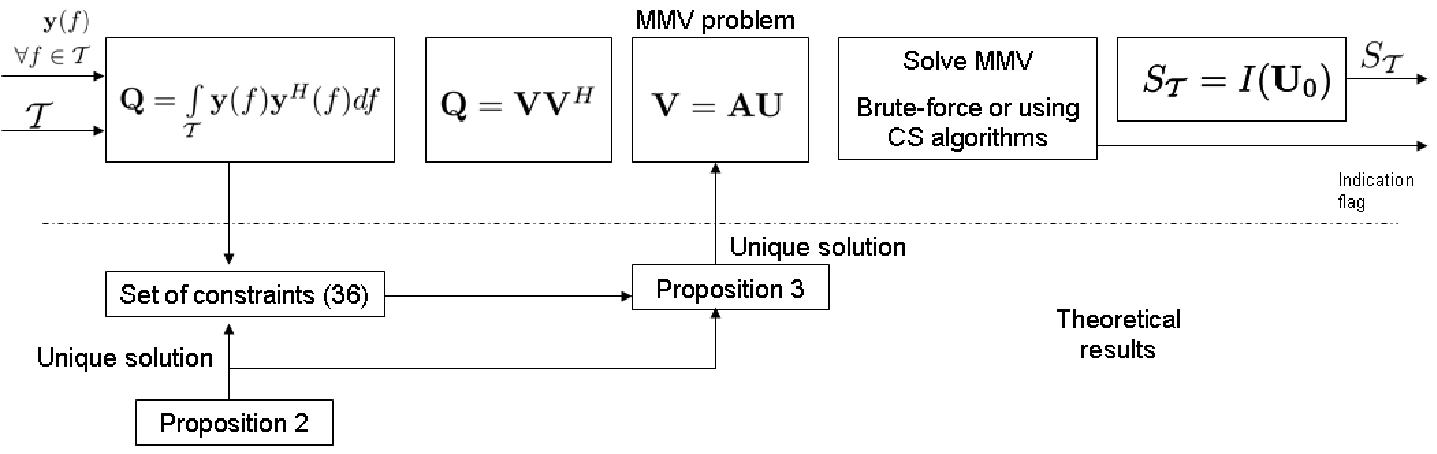}
\caption{Continuous to finite block (CTF). This block determines
the diversity set $\m(S)_\m(T)$ of a given interval $\m(T)$.}
\label{CTF}
\end{figure*}
\fi

In the CS literature, the linear system of (\ref{VAU}) is referred
to as an MMV system. Theoretical results regarding the sparsest
solution matrix of an MMV system are given in \cite{Chen}. Finding
the solution matrix $\b(U)_0$ is known to be NP-hard \cite{Davis}.
Several sub-optimal efficient algorithms for finding $\b(U)_0$ are
given in \cite{Cotter}. Some of them can indicate a success
recovery of $\b(U)_0$. We explain which class of algorithms has
this property in Section~\ref{SecSBR4}.

%%%%%%%%%%%%%%%%%%%%%%%%%%%%%%%%%%%%%%%%%%%%%%%%%%%%%%%%%%
\section{SBR algorithms}\label{SecSBRAlg}

The theoretical results developed in the previous section are now
used in order to construct the diversity set $S$ which enables the
recovery of $x(t)$ via
(\ref{ReconstructXf})-(\ref{ReconstructXfSc}).

We begin by defining a class $\m(A)$ of signals. The SBR4
algorithm is then presented and is proved to guarantee perfect
reconstruction for signals in $\m(A)$. We then show that in order
to ensure that $\m(M)\subseteq\m(A)$ the sampling rate must be at
least $4NB$, which is twice the minimal rate stated in
Theorem~\ref{BlindLandau}. To improve on this result, we define a
class $\m(B)$ of signals, and introduce a conceptual method to
perfectly reconstruct this class. The SBR2 algorithm is developed
so that it ensures exact recovery for a subset of $\m(B)$. We then
prove that $\m(M)$ is contained in this subset even for sampling
at the minimal rate. However, the computational complexity of SBR2
is higher than that of SBR4. Since universal patterns lead to the
largest sets $\m(A)$ and $\m(B)$, we assume throughout this
section that universal patterns are used, which results in
$\krank(\b(A))=p$.

%%%%%%%%%%%%%%%%%%%%%%%%%%%%%%%%%%%%%%%%%%%%%%%%%%%%%%%%%%
\subsection{The SBR4 algorithm}\label{SecSBR4}

Define the class $\m(A)_K$ of signals
\begin{equation}
\m(A)_K = \{\supp X(f)\subseteq\m(F)\,\textrm{ and }\,|S|\leq K\},
\end{equation}
with $S$ given by (\ref{setS}). Let $\m(T)=\m(F)_0$, and observe
that a multi-coset system with $p \geq 2K$ ensures that all the
conditions of Proposition~\ref{TheoremFinite} are valid for every
$x(t)\in\m(A)_K$. Thus, applying the CTF block on $\m(T)=\m(F)_0$
results in a unique sparsest solution $\b(U)_0$, with
$S=I(\b(U)_0)$. The reconstruction of the signal is then carried
out by (\ref{ReconstructXf})-(\ref{ReconstructXfSc}). We note that
(\ref{CondS}) is valid as it represents the class $\m(A)_p$ that
contains $\m(A)_K$ for $p\geq 2K$.
%Alternatively, we could have set
%$p=2K$ for some $K\in\mathbb{Z}$ and claim that $\m(A)_K$ is
%guaranteed to be perfectly reconstructed by the CTF block on
%$\m(T)=\m(F)_0$.

Algorithm~\ref{AlgSBR4}, named SBR4, follows the steps of the CTF
block in Fig. \ref{CTF} to recover the diversity set $S$ from
$\b(y)(f)$, for any $x(t)\in\m(A)_K$. The algorithm also outputs
an indication flag which we discuss later on.
%\begin{algorithm}[h]
%\caption{SBR4} \alginout{$\b(y)(f)$, Assume:
%$\krank(\b(A))=p$}{the set $S$, flag}
%\begin{algtab}
%Set $\m(T)=\m(F)_0$\\
%Compute the matrix $\b(Q)$ by (\ref{MatQ}).\\
%Decompose $\b(Q)=\b(V)\b(V)^H$ according to Proposition~\ref{PropMMV}\\
%Solve the MMV system $\b(V)=\b(A)\b(U)$ for the sparsest solution $\b(U)_0$ \alglabel{SBR4MMV}\\
%$S = I(\b(U)_0)$\\
%flag = $\{|S|\leq \frac{p}{2}\}$\alglabel{checkIndicationStep}\\
%\algreturn $S$, flag
%\end{algtab}
%\end{algorithm}
%
%% This is SBR4 algorithm

\begin{algorithm}[h]
\caption{SBR4}\label{AlgSBR4}
\begin{algorithmic}[1]
\REQUIRE $\b(y)(f)$, \textbf{Assume:} $\krank(\b(A))=p$

\ENSURE the set $S$, flag

\STATE Set $\m(T)=\m(F)_0$

\STATE Compute the matrix $\b(Q)$ by (\ref{MatQ})

\STATE Decompose $\b(Q)=\b(V)\b(V)^H$ according to
Proposition~\ref{PropMMV}

\STATE Solve the MMV system $\b(V)=\b(A)\b(U)$ for the sparsest
solution $\b(U)_0$ \label{SBR4MMV}

\STATE $S = I(\b(U)_0)$

\STATE flag = $\{|S|\leq \frac{p}{2}\}$\label{checkIndicationStep}

\RETURN $S$, flag

\end{algorithmic}
\end{algorithm}

The SBR4 algorithm guarantees perfect reconstruction of signals in
$\m(A)_K$ from samples at twice the Landau rate, which is also the
lower bound stated in Theorem~\ref{BlindLandau}. To see this,
observe that (\ref{setS}) implies that every $x(t)\in\m(A)_K$ must
satisfy
\begin{equation}\label{LandauConv}
\lambda(\supp X(f))\leq \frac{K}{LT}.
\end{equation}
Although $\m(A)_K$ is not a subspace, we use (\ref{LandauConv}) to
say that the Landau rate for $\m(A)_K$ is $K/LT$ as it contains
subspaces whose widest support is $K/LT$. As we proved, $p\geq 2K$
ensures perfect reconstruction for $\m(A)_K$. Substituting the
smallest possible value $p=2K$ into (\ref{MC_rate}) results in an
average sampling rate of $2K/LT$.

It is easy to see that flag is equal to 1 for every signal in
$\m(A)_K$. However, when a sub-optimal algorithm is used to solve
the MMV in step~\ref{SBR4MMV} we cannot guarantee a correct
solution $\b(U)_0$. Thus, flag=0 indicates that the particular MMV
method we used failed, and we may try a different MMV approach.

Existing algorithms for MMV systems can be classified into two
groups. The first group contains algorithms that seek the sparsest
solution matrix $\b(U)_0$, e.g. Basis Pursuit \cite{BPref} or
Matching Pursuit \cite{MPref} with a termination criterion based
on the residual. The other contains methods that approximate a
sparse solution according to user specification, e.g. Matching
Pursuit with a predetermined number of iterations. Using a
technique from the latter group neutralizes the indication flag as
the approximation is always sparse. Therefore, this set of
algorithms should be avoided if an indication is desired.

An important advantage of algorithm SBR4 is that the matrix
$\b(Q)$ can be computed in the time domain from the known
sequences $x_{c_i}[n],\,\, 1\leq i\leq p$. The computation
involves a set of digital filters that do not depend on the signal
and thus can be designed in advance. The exact details are given
in Appendix~\ref{SecMatrixQTime}.

The drawback of the set $\m(A)_K$, is that typically we do not
know the value of $K$. Moreover, even if $K$ is known, then
usually we do not know in advance whether $x(t)\in\m(A)_K$ as
$\m(A)_K$ does not characterize the signals according to the
number of bands and their widths. Therefore, we would like to
determine conditions that ensure $\m(M)\subseteq\m(A)_K$.
Proposition~\ref{lemmaSizeS} shows that for $x(t)\in\m(M)$ the set
$S$ satisfies $|S|\leq 2N$ if $L\leq 1/BT$. Thus, under this
condition on $L$ we have $\m(M)\subseteq\m(A)_{2N}$, which in turn
implies $p=4N$ as a minimal value for $p$. Consequently, SBR4
guarantees perfect reconstruction for $\m(M)$ under the
restrictions $L\leq 1/BT$ and $p\geq 4N$. However, the Landau rate
for $\m(M)$ is $NB$, while $p=4N$ implies a minimal sampling rate
of $4NB$. Indeed, substituting $p=4N$ and $L\leq 1/BT$ into
(\ref{MC_rate}) we have
\begin{equation}\label{SBR4rate}
\frac{p}{LT}\geq\frac{4N}{T\frac{1}{BT}}=4NB.
\end{equation}
In contrast, it follows from Theorem~\ref{SBR-Uniqueness} that
$p\geq 2N$ is sufficient for uniqueness of the solution. The
reason for the factor of two in the sampling rate is that
$\b(x)(f)$ is $N$-sparse for each specific $f$; however, when
combining the frequencies, the maximal size of $S$ is $2N$. The
SBR2 algorithm, developed in the next section, capitalizes on this
difference to regain the factor of two in the sampling rate, and
thus achieves the minimal rate, at the expense of a more
complicated reconstruction method.

%%%%%%%%%%%%%%%%%%%%%%%%%%%%%%%%%%%%%%%%%%%%%%%%%%%%%%%%%%
\subsection{The SBR2 algorithm}\label{SecSBR2}

We now would like to reduce the sampling rate required for signals
of $\m(M)$ to its minimum, i.e. twice the Landau rate. To this
end, we introduce a set $\m(B)_K$ for which SBR2 guarantees
perfect reconstruction, and then prove that
$\m(M)\subseteq\m(B)_N$ if $p\geq 2N$.

Consider a partition of $\m(F)_0$ into $M$ consecutive intervals
defined by
\begin{equation*}
0 = \bar{d}_1 < \bar{d}_2 < \cdots < \bar{d}_{M+1} = \frac{1}{LT}.
\end{equation*}
For a given partition set $\bar{D}=\{\bar{d}_i\}$ we define the
set of signals
\begin{equation*}
\m(B)_{K,\bar{D}} =\{\supp X(f)\subseteq\m(F) \textrm{ and
}|S_{[\bar{d}_i,\bar{d}_{i+1}]}|\leq K,\, 1\leq i \leq M\}.
\end{equation*}
Clearly, if $p\geq 2K$ then we can perfectly reconstruct every
$x(t)\in\m(B)_{K,\bar{D}}$ by applying the CTF block to each of
the intervals $[\bar{d}_i,\bar{d}_{i+1}]$. We now define the set
$\m(B)_K$ as
\begin{equation}
\m(B)_K = \bigcup_{\bar{D}} \m(B)_{K,\bar{D}},
\end{equation}
which is the union of $\m(B)_{K,\bar{D}}$ over all choices of
partition sets $\bar{D}$ and integers $M$. Note that neither
$\m(B)_K$ nor $\m(B)_{K,\bar{D}}$ is a subspace. If we are able to
find a partition $\bar{D}$ such that $x(t) \in \m(B)_{K,\bar{D}}$,
then $x(t)$ can be perfectly reconstructed using $p\geq 2K$. Since
the Landau rate for $\m(B)_K$ is $K/LT$, this approach requires
the minimal sampling rate\footnotemark. \footnotetext{under the
convention discussed for $\m(A)_K$.}

The following proposition shows that if the parameters are chosen
properly, then $\m(M)\subseteq\m(B)_N$. Thus, $p\geq 2N$ and a
method to find $\bar{D}$ of $x(t)$ is sufficient for perfect
reconstruction of $x(t)\in\m(M)$.

\begin{proposition}\label{PropSBR2}
If $L,p,C$ are selected according to Theorem~\ref{SBR-Uniqueness}
then $\m(M)\subseteq\m(B)_N$.
\end{proposition}
\begin{proof}
In the proof of Theorem~\ref{SBR-Uniqueness} we showed that under
the conditions of the theorem, $\b(x)(f)$ is $N$-sparse for every
$f\in\m(F)_0$. The proof of the proposition then follows from the
following lemma \cite{Bresler00}:
\begin{lemma}\label{LemmaBresler}
If $x(t)$ is a multi-band signal with $N$ bands sampled by a
multi-coset system then there exists a partition set
$\bar{D}=\{\bar{d}_i\}$ with $M=2N+1$ intervals such that
$I(\b(x)(f))$ is a constant set over the interval
$[\bar{d}_i,\bar{d}_{i+1}]$ for $1\leq i \leq M$.
\end{lemma}
Lemma~\ref{LemmaBresler} implies that
$|S_{[\bar{d}_i,\bar{d}_{i+1}]}|\leq N$ for every $1\leq i \leq
M=2N+1$ which means that $x(t)\in\m(B)_{N,\bar{D}}$.
\end{proof}

So far we showed that $\m(M)\subseteq\m(B)_N$, however to recover
$x(t)$ we need a method to find $\bar{D}$ in practice;
Lemma~\ref{LemmaBresler} only ensures its existence. Given the
data $\b(y)(f)$, our strategy is aimed at finding any partition
set $D$ such that
\begin{equation}\label{unionS}
\hat{S}=\bigcup_{i=0}^{|D|-1} S_{[d_i,d_{i+1}]}
\end{equation}
is equal to $S$, and such that $|S_{[d_i,d_{i+1}]}|\leq K$ for
every $1\leq i \leq M$. As long as (\ref{CondS}) holds, once we
find $S$ the solution is exactly recovered via
(\ref{ReconstructXf})-(\ref{ReconstructXfSc}). To find $S$, we
apply the CTF block on each interval $[d_i,d_{i+1}]$. If $p\geq
2K$, then the conditions of Proposition~\ref{TheoremFinite} are
valid, a unique solution is guaranteed for each interval. Since
for $p=2K$ (\ref{CondS}) is valid for $\m(A)_{2K}$, our method
guarantees perfect reconstruction of signals in
$\m(B)_K\cap\m(A)_{2K}$. As always, using a universal pattern
makes the set of signals $\m(B)_K\cap\m(A)_{2K}$ the largest.
Since the Landau rate for $\m(B)_K\cap\m(A)_{2K}$ is $K/LT$ this
approach allows for the minimal sampling rate when $p=2K$.

In order to find $D$ we suggest a bi-section process on $\m(F)_0$.
We initialize $\m(T)=\m(F)_0$ and seek $S_\m(T)$. If $S_\m(T)$
does not satisfy some condition explained below, then we halve
$\m(T)$ into $\m(T)_1$ and $\m(T)_2$ and determine $S_{\m(T)_1}$
and $S_{\m(T)_2}$. The bi-section process is repeated several
times until the conditions are met, or until it reaches an
interval width of no more than $\epsilon$. The set $\hat{S}$ is
then determined according to (\ref{unionS}).

We now describe the conditions for which a given $\m(T)\subseteq
\m(F)_0$ is halved. The matrix $\b(Z)_0$ of (\ref{MatZ}) satisfies
the constraints (\ref{CondZCon1})-(\ref{CondZCon2}). Since
$x(t)\in\m(A)_{2K}$ and $p\geq 2K$ (\ref{CondZCon3}) is also
valid. However, the last constraint (\ref{CondZCon4}) of
Proposition~\ref{TheoremFinite} is not guaranteed as it requires a
stronger condition $|S_\m(T)|\leq K=p/2$. Note that this condition
is satisfied immediately if $D=\bar{D}$ since $x(t)\in\m(B)_K$. We
suggest to approximate the value $S_\m(T)=|I(\b(Z)_0)|$ by
$\rank(\b(Q))$, and solve the MMV system for the sparsest solution
only if $\rank(\b(Q))\leq p/2$. This approximation is motivated by
the fact that for any $\b(Z)\succeq 0$ it is true that
$\rank(\b(Z))\leq |I(\b(Z))|$. From
Proposition~\ref{TheoremFinite} we have that
$\rank(\b(Z)_0)=\rank(\b(Q))$ which results in
\begin{equation}\label{estimateIZ}
\rank(\b(Q))\leq |I(\b(Z))|.
\end{equation}
However, only special multi-band signals result in strict
inequality in (\ref{estimateIZ}). Therefore, an interval $\m(T)$
that produces $\rank(\b(Q))>p/2$ is halved. Otherwise, we apply
the CTF block for this $\m(T)$ assuming that (\ref{estimateIZ})
holds with equality. As in SBR4 the flag indicates a correct
solution for $x(t)\in\m(B)_K\cap\m(A)_{2K}$. Therefore, if the
flag is 0 we halve $\m(T)$. These reconstruction steps are
detailed in Algorithm~\ref{AlgSBR2}, named SBR2.

%\begin{algorithm}[h]
%\caption{SBR2\label{SBR2}} \alginout{$\m(T)$, Initialize:
%$\m(T)=\m(F)_0$, Assume: $\krank(\b(A))=p$}{a set $\hat{S}$}
%\begin{algtab}
%\algifthen{$\lambda(\m(T))\leq\epsilon\;$\alglabel{StepTerminate}}{\algreturn
%$\hat{S} = \{\}$} Calculate the matrix
%\begin{equation*}
%\b(Q) = \int\limits_{f\in \m(T)} \b(y)(f)\b(y)^H(f) df
%\end{equation*}\alglabel{StepQ}\\
%\algif{$\rank(\b(Q))\leq \frac{p}{2}$\alglabel{AlgStepIfQ}}
%Decompose $\b(Q)=\b(V)\b(V)^H$\\
%Solve MMV system $\b(V)=\b(A)\b(U)$\alglabel{StepMMV}\\
%$\hat{S} = I(\b(U)_0)$\\ \algelse $\hat{S}=\{\}$\\  \algend
%\algif{$(\rank(\b(Q))>\frac{p}{2})$ \algor
%$(|\hat{S}|>\frac{p}{2})$ \alglabel{StepCondBiSect}}
%split $\m(T)$ into two equal width intervals $\m(T)_1,\m(T)_2$\\
%$\hat{S}^{(1)} = $ SBR2$(\m(T)_1)$\\
%$\hat{S}^{(2)} = $ SBR2$(\m(T)_2)$\\
%$\hat{S} = \hat{S}^{(1)} \cup \hat{S}^{(2)}$ \alglabel{StepUnion} \\
%\algend \algreturn $\hat{S}$
%\end{algtab}
%\end{algorithm}
%% This is SBR2 algorithm

\begin{algorithm}[h]
\caption{SBR2}\label{AlgSBR2}
\begin{algorithmic}[1]
\REQUIRE $\m(T)$, \textbf{Initialize:} $\m(T)=\m(F)_0$,
\textbf{Assume:} $\krank(\b(A))=p$

\ENSURE a set $\hat{S}$

\IF {$\lambda(\m(T))\leq\epsilon\;$}\label{StepTerminate}

\RETURN $\hat{S} = \{\}$

\ENDIF

\STATE Compute the matrix $\b(Q)$ by (\ref{MatQ})\label{StepQ}

\IF {$\rank(\b(Q))\leq \frac{p}{2}$}\label{AlgStepIfQ}

\STATE Decompose $\b(Q)=\b(V)\b(V)^H$

\STATE Solve MMV system $\b(V)=\b(A)\b(U)$\label{StepMMV}

\STATE $\hat{S} = I(\b(U)_0)$

\ELSE

\STATE $\hat{S}=\{\}$

\ENDIF

\IF {$(\rank(\b(Q))>\frac{p}{2})$ \textbf{ or }
$(|\hat{S}|>\frac{p}{2})$}\label{StepCondBiSect}

\STATE split $\m(T)$ into two equal width intervals
$\m(T)_1,\m(T)_2$

\STATE $\hat{S}^{(1)} = $ SBR2$(\m(T)_1)$

\STATE $\hat{S}^{(2)} = $ SBR2$(\m(T)_2)$

\STATE $\hat{S} = \hat{S}^{(1)} \cup \hat{S}^{(2)}$
\label{StepUnion}

\ENDIF

\RETURN $\hat{S}$

\end{algorithmic}
\end{algorithm}

\ifuseTwoColumns
%% This is Table II of Spectrum-blind paper
%%
\begin{table*} \caption{Spectrum-blind reconstruction methods for multi-band signals}
\begin{tabular}[C]{|l|c|c|c|}   \hlx{h}
& WKS theorem & SBR4 & SBR2 \\
\hlx{vhv} Sampling method & Uniform & Multi-coset & Multi-coset \\

Fully-blind & Yes & Yes & Yes \\

\# Uniform sequences & 1 & $p$ & $p$ \\

Minimal sampling rate & Nyquist & 2 $\times$ Landau & 2 $\times$ Landau \\

Achieves lower bound of Theorem~\ref{BlindLandau} & No & Yes & Yes
\\

Reconstruction method & Ideal low pass & SBR4 & SBR2 \\

Time complexity & constant &
1 MMV system & bi-section + finite \# of MMV \\

Applicability & $\supp X(f)\subseteq\m(F)$ &
$x(t)\in\m(A)_K$ & $x(t)\in\m(B)_K\cap\m(A)_{2K}$\footnotemark\\

Indication & No & for $x(t)\in\m(A)_K$ only & No \\
 \hlx{vhs}
\end{tabular}
\label{CompareMethods}
\end{table*}
\footnotetext{except for special signals discussed in
Section~\ref{SecSBR2}.}
\fi

It is important to note that SVR2 is sub-optimal, since the final
output of the algorithm $\hat{S}$ may not be equal to $S$ even for
$x(t)\in\m(B)_K\cap\m(A)_{2K}$. One reason this can happen is if
strict inequality holds in (\ref{estimateIZ}) for some interval
$\m(T)$. In this scenario step~\ref{StepMMV} is executed even
though $\b(Z)_0$ does not satisfy (\ref{CondZCon4}). For example,
a signal $x(t)$ with two equal width bands $[a_1,a_1+W]$ and
$[a_2,a_2+W]$ such that
\begin{equation}
\left\lfloor\frac{a_1}{LT}\right\rfloor =
\left\lfloor\frac{a_2}{LT}\right\rfloor = \gamma
\end{equation}
and $\gamma+W \in \m(F)_0$. If $x(t)$ also satisfies
\begin{equation}
X(f-a_1)=X(f-a_2),\quad\forall f\in[0,W],
\end{equation}
then it can be verified that $|I(\b(Z)_0)|=2$ while
$\rank(\b(Z)_0)=\rank(\b(Q))=1$ on the interval
$\m(T)=[\gamma,\gamma+W]$. This is of course a rare special case.
Another reason is a signal for which the algorithm reached the
termination step~\ref{StepTerminate} for some small enough
interval. This scenario can happen if two or more points of
$\bar{D}$ reside in an interval width of $\epsilon$. As an empty
set $\hat{S}$ is returned for this interval, the final output may
be missing some of the elements of $S$. Clearly, the value of
$\epsilon$ influences the amount of cases of this type. We note
that since we do not rely on $D=\bar{D}$ the missing values are
typically recovered from other intervals. Thus, both of these
sources of error are very uncommon.

The most common case in which SBR2 can fail is due to the use of
sub-optimal algorithms to find $\b(U)_0$; this issue also occurs
in SBR4. As explained before, we assume that flag=0 means an
incorrect solution and halves the interval $\m(T)$. An interesting
behavior of MMV methods is that even if $\b(U)_0$ cannot be found
for $\m(T)$, the algorithm may still find a sparse solution for
each of its subsections. Thus, the indication flag is also a way
to partially overcome the practical limitations of MMV techniques.
Note that the indication property is crucial for SBR2 as it helps
to refine the partition $D$ and reduce the sub-optimality
resulting from the MMV algorithm.

\ifuseTwoColumns
%% This is Table III of Spectrum-blind paper
%%
\begin{table} \caption{Comparison of SBR4 and SBR2 for signals in $\m(M)$}
\begin{tabular}[C]{|l|c|c|}   \hlx{h}
& SBR4 & SBR2 \\
\hlx{vhv}
\# Uniform sequences & $p\geq 4N$ & $p\geq 2N$ \\

Minimal rate & 4 $\times$ Landau & 2 $\times$ Landau \\

Lower bound of Th.~\ref{BlindLandau} & No & Yes
\\

Parameter selection & Theorem~\ref{SBR-Uniqueness}, $p\geq 4N$ &
Theorem~\ref{SBR-Uniqueness}\\

Perfect reconstruction & Yes & Yes\footnotemark[3]
\\

Indication & Yes & No \\
 \hlx{vhs}
\end{tabular}
\label{CompareMethodsM}
\end{table}
\fi

We point out that Proposition~\ref{PropSBR2} shows that
$\m(M)\subseteq\m(B)_N$. We also have that
$\m(M)\subseteq\m(A)_{2N}$ from Proposition~\ref{lemmaSizeS},
which motivates our approach. The SBR2 algorithm itself does not
impose any additional limitations on $L,p,C$ other than those of
Theorem~\ref{SBR-Uniqueness} required to ensure the uniqueness of
the solution. Therefore, theoretically, perfect reconstruction for
$\m(M)$ is guaranteed if the samples are acquired at the minimal
rate, with the exception of the special cases discussed before.

The complexity of SBR2 is dictated by the number of iterations of
the bi-section process, which is also affected by the behavior of
the MMV algorithm that is used. Numerical experiments in Section
\ref{SecNumerical} show that empirically SBR2 converges
sufficiently fast for practical usage.

\ifuseTwoColumns%% This figure is the SBR scheme
\begin{figure*}
\centering
\includegraphics[scale=0.6]{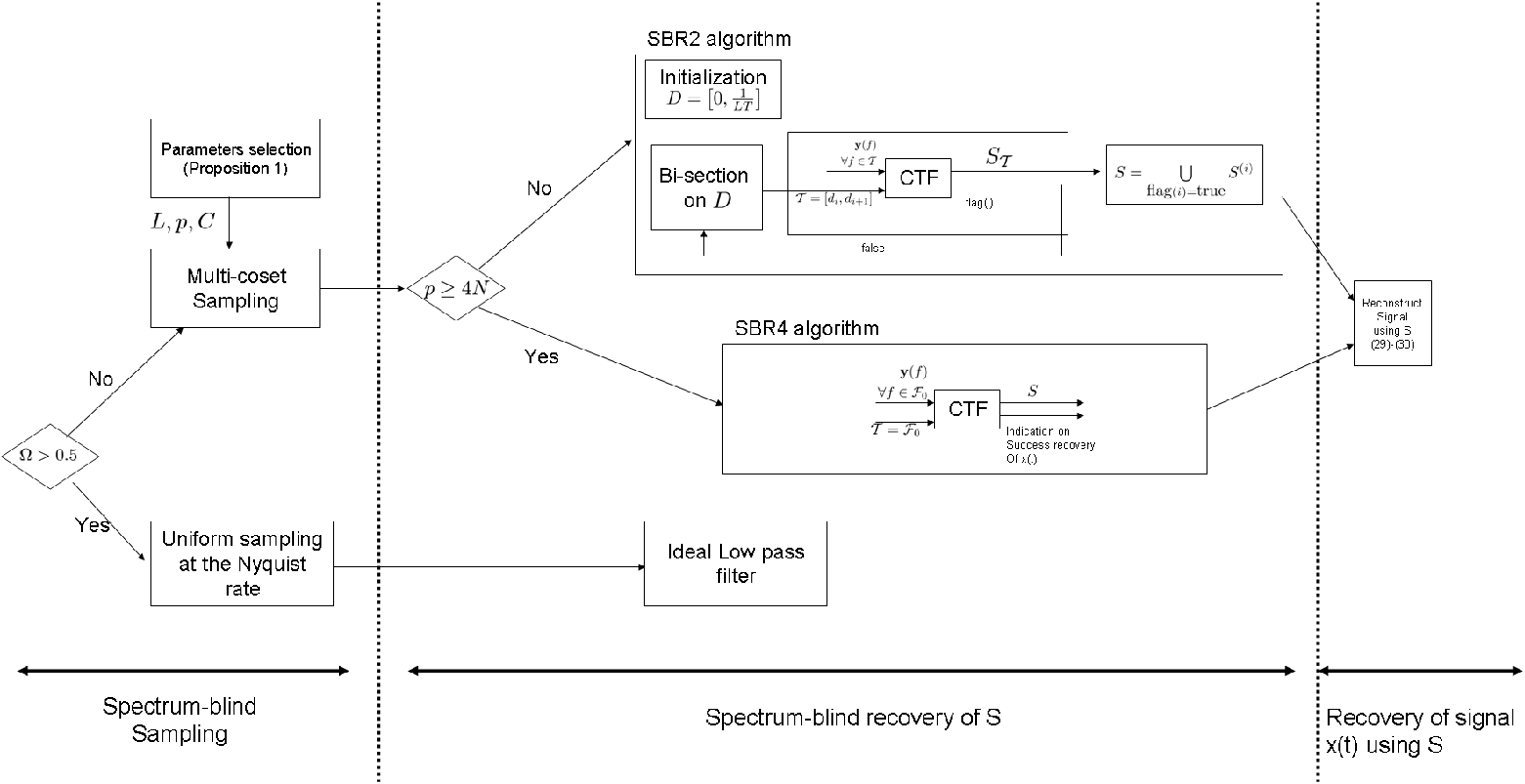}
\caption{Spectrum-blind reconstruction scheme.} \label{SBRscheme}
\end{figure*}
\fi

Finally, we emphasize that SBR2 does not provide an indication on
the success recovery of $x(t)$ even for $x(t)\in\m(M)$ since there
is no way to know in advance if $x(t)$ is a signal of the special
type that SBR2 cannot recover.

%%%%%%%%%%%%%%%%%%%%%%%%%%%%%%%%%%%%%%%%%%%%%%%%%%%%%%%%%%
\subsection{Comparison between SBR4 and SBR2}

Table~\ref{CompareMethods} compares the properties of SBR4 and
SBR2. We added the WKS theorem as it also offers spectrum-blind
reconstruction. Both SBR4 and SBR2 algorithms recover the set $S$
according to the paradigm stated in Section~\ref{SecParadigm}.
Observe that an indication property is available only for SBR4 and
only if the signals are known to lie in $\m(A)_K$. Although both
SBR4 and SBR2 can operate at the minimal sampling rate, SBR2
guarantees perfect reconstruction for a wider set of signals as
$\m(A)_K$ is a true subset of $\m(B)_K\cap\m(A)_{2K}$.

\ifuseTwoColumns\else
%% This is Table II of Spectrum-blind paper
%%
\begin{table*} \caption{Spectrum-blind reconstruction methods for multi-band signals}
\begin{tabular}[C]{|l|c|c|c|}   \hlx{h}
& WKS theorem & SBR4 & SBR2 \\
\hlx{vhv} Sampling method & Uniform & Multi-coset & Multi-coset \\

Fully-blind & Yes & Yes & Yes \\

\# Uniform sequences & 1 & $p$ & $p$ \\

Minimal sampling rate & Nyquist & 2 $\times$ Landau & 2 $\times$ Landau \\

Achieves lower bound of Theorem~\ref{BlindLandau} & No & Yes & Yes
\\

Reconstruction method & Ideal low pass & SBR4 & SBR2 \\

Time complexity & constant &
1 MMV system & bi-section + finite \# of MMV \\

Applicability & $\supp X(f)\subseteq\m(F)$ &
$x(t)\in\m(A)_K$ & $x(t)\in\m(B)_K\cap\m(A)_{2K}$\footnotemark\\

Indication & No & for $x(t)\in\m(A)_K$ only & No \\
 \hlx{vhs}
\end{tabular}
\label{CompareMethods}
\end{table*}
\footnotetext{except for special signals discussed in
Section~\ref{SecSBR2}.}
\fi

Considering signals from $\m(M)$ we have to restrict the parameter
selection. The specific behavior of SBR4 and SBR2 for this
scenario is compared in Table~\ref{CompareMethodsM}. In
particular, SBR4 requires twice the minimal rate.

\ifuseTwoColumns\else
%% This is Table III of Spectrum-blind paper
%%
\begin{table} \caption{Comparison of SBR4 and SBR2 for signals in $\m(M)$}
\begin{tabular}[C]{|l|c|c|}   \hlx{h}
& SBR4 & SBR2 \\
\hlx{vhv}
\# Uniform sequences & $p\geq 4N$ & $p\geq 2N$ \\

Minimal rate & 4 $\times$ Landau & 2 $\times$ Landau \\

Lower bound of Th.~\ref{BlindLandau} & No & Yes
\\

Parameter selection & Theorem~\ref{SBR-Uniqueness}, $p\geq 4N$ &
Theorem~\ref{SBR-Uniqueness}\\

Perfect reconstruction & Yes & Yes\footnotemark[3]
\\

Indication & Yes & No \\
 \hlx{vhs}
\end{tabular}
\label{CompareMethodsM}
\end{table}
\fi

In the tables, perfect reconstruction refers to reconstruction
with a brute-force MMV method that finds the correct solution. In
practice, sub-optimal MMV algorithms may result in failure of
recovery even when the other requirements are met. The indication
flag is intended to discover these cases.

The entire reconstruction scheme is presented in
Fig.~\ref{SBRscheme}. The scheme together with the tables allow
for a wise decision on the particular implementation of the
system. Clearly, for $\Omega>0.5$ it should be preferred to sample
at the Nyquist rate and to reconstruct with an ideal low pass
filter. For $\Omega\leq 0.5$ we have to choose between SBR4 and
SBR2 according to our prior on the signal. Typically, it is
natural to assume $x(t)\in\m(M)$ for some values of $N$ and $B$
and derive the required parameter selection according to
Table~\ref{CompareMethodsM}. It is obvious that if $p\geq 4N$ is
used then SBR4 should be preferred since it is less complicated
than SBR2.

\ifuseTwoColumns\else%% This figure is the SBR scheme
\begin{figure*}
\centering
\includegraphics[scale=0.6]{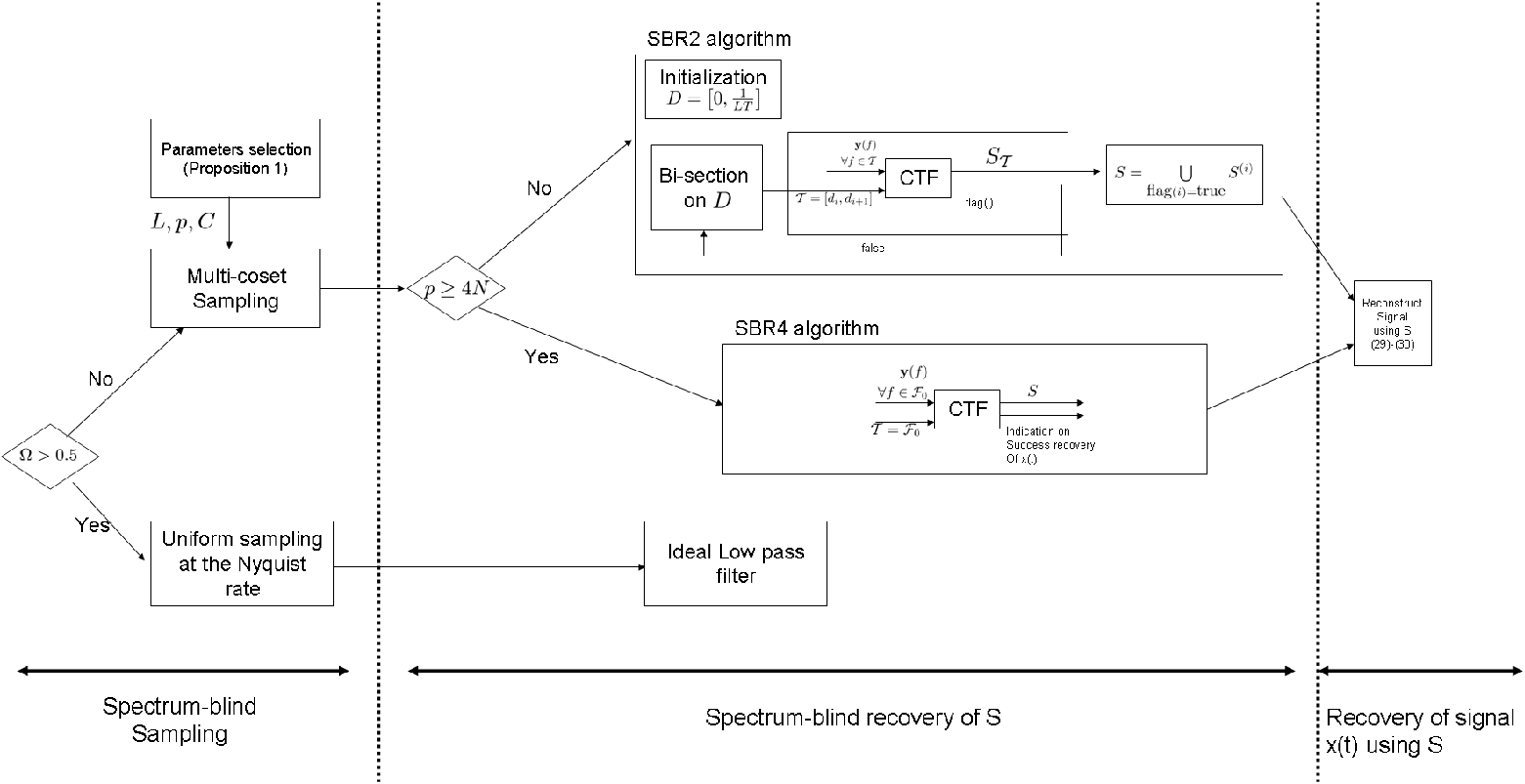}
\caption{Spectrum-blind reconstruction scheme.} \label{SBRscheme}
\end{figure*}
\fi

The trade-off presented here between complexity and sampling rate
also exists in the known-spectrum reconstruction of
\cite{Bresler00}. Sampling at the minimal rate of Landau requires
a reconstruction that consists of piecewise constant filters. The
number of pieces and the reconstruction complexity grow with $L$.
This complexity can be prevented by doubling the value of $p$
which also doubles the average sampling rate according to
(\ref{MC_rate}). Then,
(\ref{ReconstructXf})-(\ref{ReconstructXfSc}) are used to
reconstruct the signal by only one inversion of a known matrix
\cite{Bresler96_1D}.

%%%%%%%%%%%%%%%%%%%%%%%%%%%%%%%%%%%%%%%%%%%%%%%%%%%%%%%%%%
\section{Numerical experiments}\label{SecNumerical}

We now provide several experiments demonstrating the
reconstruction using algorithms SBR4 and SBR2 for signals from
$\m(M)$. We also provide an example in which the signals do not
lie in the class $\m(M)$ but in the larger set implied by
$\m(A)_K$ for SBR4 and by $\m(B)_K\cap\m(A)_{2K}$ for SBR2.

%%%%%%%%%%%%%%%%%%%%%%%%%%%%%%%%%%%%%%%%%%%%%%%%%%%%%%%%%%
\subsection{Setup}\label{SecNumSetup}

The setup described hereafter is used as a basis for all the
experiments.

Consider an example of the class $\m(M)$ with $\m(F)=[0,20\textrm{
GHz}], N=4$ and $B=100$ MHz. In order to test the algorithms 1000
test cases from this class were generated randomly according to
the following steps:
\begin{enumerate}
\item draw $\{a_i\}_{i=1}^N$ uniformly at random from
$[0,20\textrm {GHz}-B]$.\label{SimSetupStep1} \item set
$b_i=a_i+B$ for $1\leq i \leq N$, and ensure that the bands do not
overlap. \label{SimSetupStep2} \item Generate $X(f)$ by
\begin{numcases}{X(f)=}
\nonumber \alpha(f)\left(S_R(f) + j S_I(f)\right), & $f\in\bigcup\limits_{i=1}^N [a_i,b_i]$  \\
\nonumber 0, & otherwise.
\end{numcases}
For every $f$ the values of $S_R(f)$ and $S_I(f)$ are drawn
independently from a normal distribution with zero mean and unit
variance. The function $\alpha(f)$ is constant in each band, and
is chosen such that the band energy is equal to $e_i$ where $e_i$
is selected uniformly from [1,5].
\end{enumerate}
The Landau rate for each of the signals is $NB=400$ MHz, and thus
the minimal rate requirement for blind reconstruction is $800$ MHz
due to Theorem \ref{BlindLandau}.

Several multi-coset systems are considered with the following
parameters. The value $L$ is common in all the systems. The value
of $p$ is varied from $p=N=4$ to $p=8N=32$ representing 29
different systems. A universal pattern $C$ is constructed by
choosing prime $L$, since according to \cite{MishaliUniv} this
ensures that every sampling pattern is universal.

An experiment is conducted by sampling the signals using each of
the multi-coset systems. Each of these combinations is used as an
input to both SBR4 and SBR2 algorithms. We selected the
Multi-Orthogonal Matching Pursuit (M-OMP) method \cite{Cotter} to
solve the MMV systems for the sparsest solution. The empirical
success rate of each algorithm is calculated as the ratio of
simulations in which the recovered set $S$ is correct.

%%%%%%%%%%%%%%%%%%%%%%%%%%%%%%%%%%%%%%%%%%%%%%%%%%%%%%%%%%
\subsection{Sampling rate and practical limitations}\label{exp1}

We begin by selecting the largest possible value of prime $L$
satisfying (\ref{LBT}):
\begin{equation}
L = 199 \leq \frac{1}{BT}=200.
\end{equation}
Thus, the minimal rate requirement holds only for $p\geq 2N$.
Specifically, for $p=2N$ the sampling rate is $p/LT=804$ MHz.
Observe that a non-prime $L=200$ would give the minimal rate
exactly. This setting is discussed later on.

\begin{figure}
\centering
\includegraphics[scale=0.7]{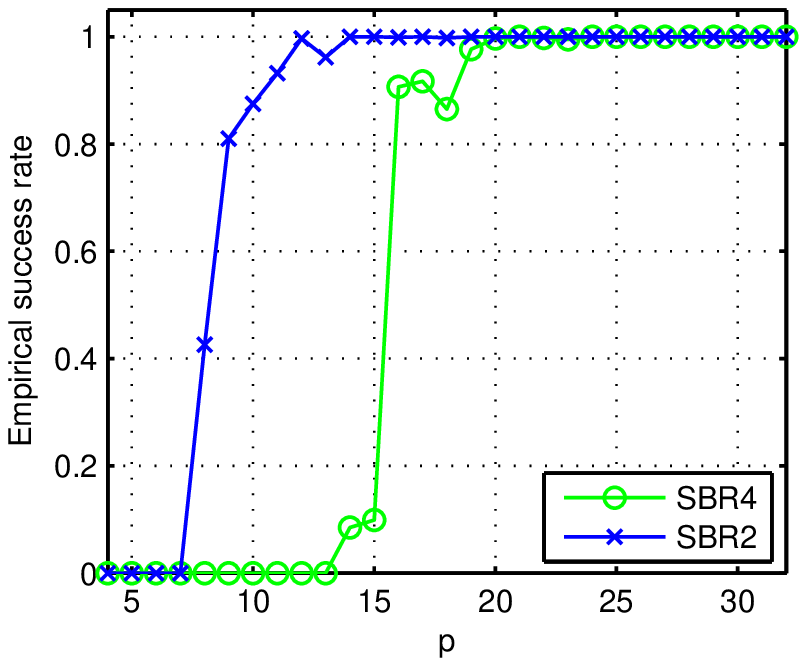}
\caption{Performance of SBR algorithms with $L=199$.} \label{Fig1}
\end{figure}

Fig.~\ref{Fig1} depicts the empirical success rate with
$L=199,N=4$ as a function of $p$. It is evident that for $p<2N$
the set $S$ could not be recovered by neither of the algorithms
since the sampling rate is below the bound of
Theorem~\ref{BlindLandau}. As expected, SBR2 outperforms SBR4 as
it achieves the same empirical success rate for a lower average
sampling rate. It is also seen that for $p=4N$ the sampling rate
is slightly more than four times the Landau rate. Indeed,
algorithm SBR4 maintains a high recovery rate for this value of
$p$. The usage of SBR2 with M-OMP maintains a high recovery rate
for $p/N=2.6$, which is more than the minimal rate. Other MMV
algorithms may be used to improve this result, however we used
only M-OMP as it is simple and fast.

\begin{figure}
\centering
\includegraphics[scale=0.7]{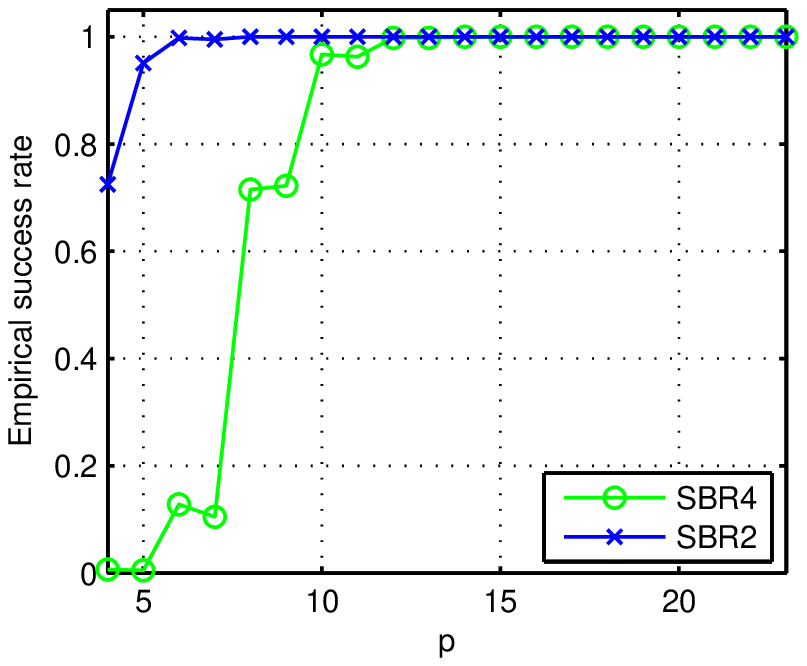}
\caption{Performance of SBR algorithms with $L=23$.} \label{Fig2}
\end{figure}

We next consider a scenario with $L=23$, which  clearly satisfies
(\ref{LBT}). Here, for $p=N=4$ we have a sampling rate of 3.4 GHz
which is much higher than the minimal requirement. This selection
of $L$ represents a practical desire to satisfy the minimal rate
requirement with a reduced value of $p$, since realizing the
multi-coset sampling requires $p$ analog-to-digital devices.
Fig.~\ref{Fig2} presents the empirical recovery rate in this case.
Note that Table \ref{CompareMethodsM} shows that in order to
guarantee perfect reconstruction for $\m(M)$ we need $p\geq 4N$
for SBR4, and $p\geq 2N$ for SBR2. However, these conditions are
only sufficient. Indeed, it is evident from Fig.~\ref{Fig2} that
both algorithms reach a satisfactory recovery rate for lower
values of $p$ .

In Table~\ref{TableRunTimes}, we tabulate the average run time of
one case out of the 1000 tested. Our experiments were conducted on
an ordinary PC desktop with an Intel CPU running at 2.4GHz and
512MB memory RAM. We used Matlab version 7 to encode and execute
the algorithms. Note that for $L=199,p=2N$ we encountered a
significant increase in SBR2 runtime. The reason is that the
average sampling rate is very close to the minimal possible, thus
the recursion depth of the algorithm grows as it is harder to find
a suitable partition set $D$. For $p=4N$ the runtime dramatically
improves, however in this case SBR4 may be preferred due to the
advantages that appear in Table~\ref{CompareMethodsM}. It can be
seen that for $L=23$ the average runtime is low for both
algorithms. This scenario represents a case that the value of
$|S|$ is very low compared to $2N$, and thus it is easier to find
a partition set $D$. Moreover, M-OMP becomes faster as the
solution is sparser.

\begin{table} \caption{Average run time of SBR4 and SBR2 with MOMP (msec) }
\begin{tabular}[C]{|l|c|c|c|c|}   \hlx{h}
 & \multicolumn{2}{c|}{ \T \B$L=199$} & \multicolumn{2}{c|}{$L=23$}\\
 \cline{2-5}
 & \T \B SBR4 & SBR2  &  SBR4 & SBR2 \\    \hlx{vhv}
\T\B$p=N$ & 7 & 608 & 4.2 & 51.4\\
\T\B$p=2N$ & 16.1 & 1034 & 5.7 & 6.4\\
\T\B$p=4N$ & 21.4 & 24.8 & 6.7 & 6.7\\ \hlx{vhs}
\end{tabular}
\label{TableRunTimes}
\end{table}

%%%%%%%%%%%%%%%%%%%%%%%%%%%%%%%%%%%%%%%%%%%%%%%%%%%%%%%%%%
\subsection{Applicability}

The previous experiments demonstrated the applicability of SBR4
and SBR2 to signals that lie in $\m(M)$. We now explore the case
in which $x(t)\notin\m(M)$.

In this experiment we used the basic setup with $L=199$ but the
signals are constructed in a different way. Each one of the 1000
signals is constructed by $X(f)=\alpha(f)\left(S_R(f) + j
S_I(f)\right),\,\forall f\in\m(F)_0$. The function $\alpha(f)$
depends on the algorithm and it makes sure that $x(t)\in\m(A)_K$
for the test cases of SBR4. Similarly, $\alpha(f)$ is used to form
signals $x(t)\in\m(B)_K\cap\m(A)_{2K}$ for SBR2. The construction
of these signals depends on $L$ because of the definitions of
$\m(A)_K$ and $\m(B)_K$. We selected $K=8$ which results in a
Landau rate of $K/LT=804$ MHz in either construction. In addition,
we made sure that the signals do not lie in $\m(M)$.

\begin{figure}
\centering
\includegraphics[scale=0.7]{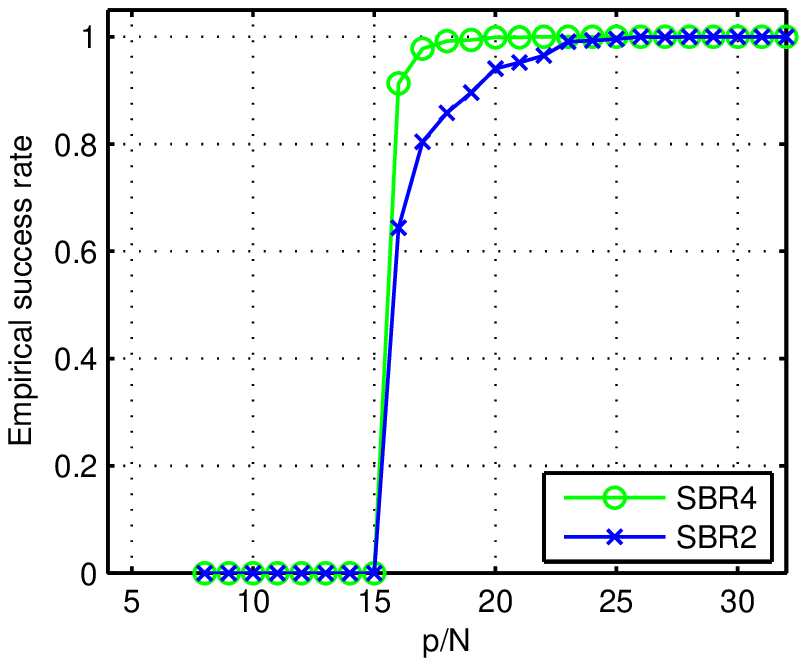}
\caption{Performance for signals $x(t)$ not in
$\m(M)$.}\label{Fig3}
\end{figure}

Fig.~\ref{Fig3} shows the empirical recovery rate of SBR4 and SBR2
in this scenario. The value $p=4N=16$ serves as a threshold for
satisfactory recovery, as the sampling rate for this value of $p$
is $p/LT=1608$ MHz, which is twice the Landau rate. It can also be
seen that SBR4 performs better than SBR2 as it does not involve a
sub-optimal stage of recovering the partition set $D$. Both
algorithms suffer from the sub-optimality techniques for MMV
systems.

Note that the signals here are synthesized so that they lie in the
relevant sets. However, for a generic signal $x(t)\notin\m(M)$
there is no way to know in advance whether it lies in one of these
sets. Moreover, there is no way to infer it from the samples,
$\b(y)(f)$. In addition, even if SBR4 is used for this signal and
it returns flag=1, there is no meaning for this indication since
the uniqueness of the solution is guaranteed only for
$x(t)\in\m(A)_K$ which cannot be ensured for a generic multi-band
signal.

%%%%%%%%%%%%%%%%%%%%%%%%%%%%%%%%%%%%%%%%%%%%%%%%%%%%%%%%%%
\subsection{Random sampling patterns}

\begin{figure}
\centering
\includegraphics[scale=0.7]{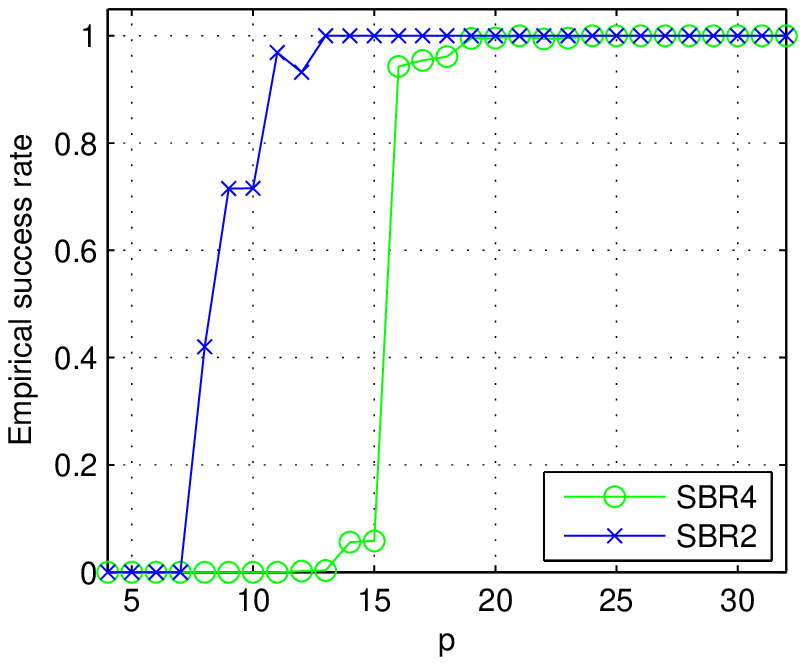}
\caption{Performance of SBR algorithms with $L=200$.} \label{Fig4}
\end{figure}

Theorem~\ref{SBR-Uniqueness} requires a universal sampling
pattern, which means finding a pattern resulting in
$\krank(\b(A))=p$. However, computing the value of $\krank(\b(A))$
requires a combinatorial process for non-prime $L$. The "bunched"
pattern $C=\{0,1,...,p-1\}$ given in \cite{Bresler00} is proved to
be universal but the matrix $\b(A)$ is not well conditioned for
this choice \cite{Bresler96}. Alternatively, it follows from the
work of Cand\`{e}s \textit{et. al.} \cite{CandesRobust} that
random sampling patterns are most likely to produce a high value
for $\krank(\b(A))$ if $L,p$ are large enough. Therefore, for
practical usage the sampling pattern can be selected randomly even
for non-prime $L$. Fig.~\ref{Fig4} presents an experiment with
$L=200$. We point out that the random selection process is carried
out only once, and the same sampling patterns are used for all the
tested signals. Comparing Figs.~\ref{Fig1} and~\ref{Fig4} it is
seen that the results are very similar although the exact value of
$\krank(\b(A))$ is unknown.

This experiment was also performed when for every $N\leq p\leq 8N$
the patterns are selected as $C=\{c_k|c_k=2k,\,0\leq k \leq
p-1\}$, which is proved in \cite{MishaliUniv} to render
$\krank(\b(A))=1$. In this case both SBR4 and SBR2 could not
recover any of the 1000 test cases. Thus, the universality of the
pattern is crucial to the success of our method.

%%%%%%%%%%%%%%%%%%%%%%%%%%%%%%%%%%%%%%%%%%%%%%%%%%%%%%%%%%
\section{Conclusions}\label{SecConclusions}

In this paper we suggested a method to reconstruct a multi-band
signal from its samples when the band locations are unknown. Our
development enables a fully spectrum-blind system where both the
sampling and the reconstruction stages do not require this
knowledge.

Our main contribution is in proving that the reconstruction
problem can be formulated as a finite dimensional problem within
the framework of compressed sensing. This result is accomplished
without any discretization. Conditions for uniqueness of the
solution and algorithms to find it were developed based on known
theoretical results and algorithms from the CS literature.

In addition, we proved a lower bound on the sampling rate that
improves on the Landau rate for the case of spectrum-blind
reconstruction. One of the algorithms we proposed indeed
approaches this minimal rate for a wide class of multi-band
signals characterized by the number of bands and their widths.

Numerical experiments demonstrated the trade off between the
average sampling rate and the empirical success rate of the
reconstruction.

%%%%%%%%%%%%%%%%%%%%%%%%%%%%%%%%%%%%%%%%%%%%%%%%%%%%%%%%%%
\appendices

%%%%%%%%%%%%%%%%%%%%%%%%%%%%%%%%%%%%%%%%%%%%%%%%%%%%%%%%%%
\section{Real-valued signals}\label{SecReal}

In order to treat real-valued signal the following definitions
replace the ones given in the paper. The class $\m(M)$ is changed
to contain all real-valued multi-band signals bandlimited to
$\m(F)=[-1/2T,1/2T]$ with no more than $N$ bands on both sides of
the spectrum, where each the band width is upper bounded by $B$ as
before. Note that $N$ is even as the Fourier transform is
conjugate symmetric for real-valued signals. The Nyquist rate
remains $1/T$ and the Landau rate is $NB$.

Repeating the calculations of \cite{Bresler00} that lead to
(\ref{EqYAX}) it can be seen that several modifications are
required as now explained. To form $\b(x)(f)$, the interval
$\m(F)$ is still divided into $L$ equal intervals. However, a
slightly different treatment is given for odd and even values of
$L$, because of the negative side of the spectrum. Define the set
of $L$ consecutive integeres
\begin{numcases}{K=}
\nonumber \left\{-\frac{L-1}{2},\cdots,\frac{L-1}{2}\right\},&odd $L$\\
\nonumber \left\{-\frac{L}{2},\cdots,\frac{L}{2}-1\right\},&even
$L$.
\end{numcases}
and redefine the interval $\m(F)_0$
\begin{numcases}{\m(F)_0 =}
\nonumber \left[-\frac{1}{2LT},\frac{1}{2LT}\right],&odd $L$\\
\nonumber \left[0,\frac{1}{LT}\right],&even $L$.
\end{numcases}
The vector $\b(x)(f)$ is now defined as
\begin{equation*}
\b(x)_i(f) = X(f+K_i/LT),\quad\forall 0\leq i\leq L-1,
\end{equation*}
The dimensions of $\b(A)$ remain $p \times L$ with $ik$ entry
\begin{eqnarray}
\b(A)_{ik} = \frac{1}{LT}\exp\left(j\frac{2\pi}{L}c_i
K_k\right),\\\nonumber \, 1\leq i \leq p,\quad\, 0\leq k\leq L-1.
\end{eqnarray}
The definition of $\b(y)(f)$ remains the same with respect to
$\m(F)_0$ defined here. The results of the paper are thus extended
to real-valued multi-band signals since (\ref{My_yAx}) is now
valid with respect to these definitions of $\b(x)(f)$, $\b(A)$,
and $\m(F)_0$.

Note that, we could have, conceptually, constructed a
complex-valued multi-band signal by taking only the positive
frequencies of the real-valued signal. The Landau rate of this
complex version is $NB/2$. Nevertheless, the information rate is
the same as each sample of a complex-valued signal is represented
by two real numbers.

%%%%%%%%%%%%%%%%%%%%%%%%%%%%%%%%%%%%%%%%%%%%%%%%%%%%%%%%%%
\section{Proof of Lemma~\ref{LemIn}}\label{ProofLemIn}

Let $r=\rank(\b(P))$. Reorder the columns of $\b(P)$ so that the
first $r$ columns are linearly independent. This operation does
not change the rank of $\b(P)$ nor the rank of $\b(A)\b(P)$.
Define
\begin{equation}
\b(P) = [\b(P)^{(1)}\,\,\; \b(P)^{(2)}],
\end{equation}
where $\b(P)^{(1)}$ contains the first $r$ columns of $\b(P)$ and
the rest are contained in $\b(P)^{(2)}$. Therefore,
\begin{equation*}
r \geq \rank(\b(A)\b(P)) = \rank(\b(A)[\b(P)^{(1)}\,\,\;
\b(P)^{(2)}]) \geq \rank(\b(A)\b(P)^{(1)}).
\end{equation*}
The inequalities result from the properties of the rank of
concatenation and of multiplication of matrices. So it is
sufficient to prove that $\b(A)\b(P)^{(1)}$ has full column rank.

Let $\alpha$ be a vector of coefficients so that $\b(A)\b(P)^{(1)}
\b(\alpha)=\b(0)$. It remains to prove that this implies
$\b(\alpha)=\b(0)$. Denote $k=|I(\b(P))|$. Since $I(\b(P)^{(1)})
\subseteq I(\b(P))=k$ the vector $\b(P)^{(1)}\b(\alpha)$ is
$k$-sparse. However, $\krank(\b(A))\geq k$ and its null space
cannot contain a $k$-sparse vector unless it is the zero vector.
Since $\b(P)^{(1)}$ contains linearly independent columns this
implies $\b(\alpha)=\b(0)$.

%%%%%%%%%%%%%%%%%%%%%%%%%%%%%%%%%%%%%%%%%%%%%%%%%%%%%%%%%%
\section{Computation of the matrix $\b(Q)$}\label{SecMatrixQTime}

The SBR4 algorithm computes the matrix $\b(Q)$ in the frequency
domain. A method to compute this matrix directly from the samples
in the time domain is now presented.

Consider the $ik$th element of $\b(Q)$ from (\ref{MatQ}):
\begin{equation}
\b(Q)_{ik} = \int_0^\frac{1}{LT} \b(y)_i(f)\b(y)_k^*(f) df.
\end{equation}
Since $\b(y)_i(f)$ is the DTFT of $x_{c_i}[n]$ we can write
$\b(Q)_{ik}$ as,
\begin{align}\label{MatrixQik}
& \b(Q)_{ik} = \int_0^\frac{1}{LT} \left(\sum_{n_i \in \mathbb{Z}}
x_{c_i}[n_i]\exp\left(-j2\pi f n_i T\right)\right)\cdot
\\ \notag &\phantom{\b(Q)_{ik} = \int_0^\frac{1}{LT} {}} \left(\sum_{n_k\in\mathbb{Z}}
x_{c_k}[n_k]\exp\left(-j2\pi f n_k T\right)\right)^* df\\
\notag &\phantom{\b(Q)_{ik} {}}= \sum_{n_i \in \mathbb{Z}}
\sum_{n_k \in \mathbb{Z}} x_{c_i}[n_i] x_{c_k}^*[n_k]
\int_0^\frac{1}{LT} \exp\left(j2\pi f (n_k-n_i) T\right) df.
\end{align}
Note that from (\ref{xci}) the sequence $x_{c_i}[n_i]$ is padded
by $L-1$ zeros between the non-zero samples. Define the sequence
without these zeros as
\begin{eqnarray}\label{xhat_ci}
\hat{x}_{c_i}[m] = x(m L T + c_i T),\quad m\in\mathbb{Z},\quad
1\leq i \leq p.
\end{eqnarray}
Then, (\ref{MatrixQik}) can be written as
\begin{align}\label{MatrixQiSeries}
&\b(Q)_{ik} = \sum_{m_i \in \mathbb{Z}} \sum_{m_k \in \mathbb{Z}}
\hat{x}_{c_i}[m_i] \hat{x}_{c_k}^*[m_k] g_{ik}[m_i-m_k]\\
\notag &\phantom{\b(Q)_{ik}{}} = \sum_{m_i \in \mathbb{Z}}
\hat{x}_{c_i}[m_i] (\hat{x}_{c_k} \ast g_{ik})[m_i],
\end{align}
where
\begin{equation}
g_{ik}[m] = \int_0^\frac{1}{LT} \exp\left(j2\pi f ( mL + (c_k-c_i)
) T\right) df,
\end{equation}
and
\begin{equation}
(\hat{x}_{c_k} \ast g_{ik})[m] = \sum_{n \in \mathbb{Z}}
\hat{x}_{c_k}^*[n] g_{ik}[m-n].
\end{equation}

If $i=k$ then $c_i=c_k$ and
\begin{equation}\label{MatrixQ_g_ik1}
g_{ii}[m] = g[m] = \frac{1}{LT}\exp(j\pi m)\sinc(m),
\end{equation}
with $\sinc(x)=\sin(\pi x)/(\pi x)$.

If $i\neq k$,
\begin{eqnarray}\label{MatrixQ_g_ik2}
g_{ik}[m]=\frac{\exp\left(j\frac{2\pi}{L}(c_k-c_i)\right)-1}{j2\pi
( mL + (c_k-c_i) ) T}.
\end{eqnarray}

The set of digital filter $g_{ik}$ can be designed immediately
after setting the parameter $L,p,C$ as these filters do not depend
on the signal.

\setlength{\arraycolsep}{0.0em} \setlength{\arraycolsep}{5pt}

\end{document}